\documentclass[lettersize,journal]{IEEEtran}
\usepackage{amsmath,amsfonts}
\usepackage{algorithmic}
\usepackage{algorithm}
\usepackage{array}
\usepackage{textcomp}
\usepackage{stfloats}
\usepackage{url}
\usepackage{verbatim}
\usepackage{graphicx}
\usepackage{subcaption}
\usepackage{caption}
\usepackage{cite}
\usepackage{bm}
\usepackage{amsthm,amsmath,amssymb}
\usepackage{mathrsfs}
\usepackage{soul,color,xcolor}
\usepackage{graphicx}
\usepackage{epstopdf}
\usepackage{stfloats}
\usepackage[labelsep=period]{caption}

\allowdisplaybreaks[4]
\def\BibTeX{{\rm B\kern-.05em{\sc i\kern-.025em b}\kern-.08em
		T\kern-.1667em\lower.7ex\hbox{E}\kern-.125emX}}
\usepackage{balance}
\begin{document}
	\title{Rotatable Antenna Enabled Secure Wireless Information and Power Transfer}
	\author{Yixuan Li, Ji Wang, \IEEEmembership{Senior Member,~IEEE},  Beixiong Zheng, \IEEEmembership{Senior Member,~IEEE}, \\Ming Zeng, \IEEEmembership{Senior Member,~IEEE}, and Xianbin Wang, \IEEEmembership{Fellow,~IEEE}
		\thanks{\textit{(Corresponding author: Ji Wang.)}}
		\thanks{Yixuan Li and Ji Wang are with the Department of Electronics and Information Engineering, Central China Normal University, Wuhan 430079, China (e-mail: yixuanli@mails.ccnu.edu.cn; jiwang@ccnu.edu.cn).
			
			Beixiong Zheng is with the School of Microelectronics, South China University of Technology, Guangzhou 511442, China (e-mail: bxzheng@scut.edu.cn).
			
			Ming Zeng is with the Department of Electrical Engineering and Computer Engineering, Universite Laval, Quebec, QC G1V 0A6, Canada (e-mail: ming.zeng@gel.ulaval.ca).
			
			Xianbin Wang is with the Department of Electrical and Computer Engineering, Western University, London, ON N6A 5B9, Canada (e-mail: xianbin.wang@uwo.ca).
	}}
	
	\maketitle
	
	\begin{abstract}
		Rotatable antenna (RA) provides additional spatial degrees of freedom by exploiting the rotational flexibility of directional antennas. This paper investigates an RA-enabled secure simultaneous wireless information and power transfer (SWIPT) system, where multiple RAs at the base station (BS) are independently rotated to support secure information transmission and wireless power transfer. We maximize the sum received power at the energy receivers (ERs) by jointly optimizing the transmit beamforming and RA rotation angles subject to quality-of-service, secrecy, and energy harvesting constraints. To solve the resulting non-convex problem, a performance-oriented SCA-based alternating optimization (AO) algorithm is developed, where the beamforming subproblem is solved via semidefinite relaxation (SDR) with deterministic rank-one recovery, while the RA rotations are optimized using SCA. In addition, a complexity-oriented differential evolution (DE)-based AO algorithm is developed as a complementary alternative with simpler implementation for RA rotation optimization. In addition, a worst-case robust beamforming extension is proposed for norm-bounded channel state information (CSI) errors. Numerical results demonstrate the effectiveness of the proposed algorithms and show that the RA-enabled design significantly outperforms conventional fixed- and isotropic-antenna schemes.
	\end{abstract}
	

	\section{Introduction}
	The unprecedented growth of the Internet-of-Things (IoT) applications has led to the deployment of a massive number of IoT devices to support a broad spectrum of human-centric applications. The massive deployment of IoT devices inevitably leads to prohibitively high operational and maintenance costs \cite{7123563,6714496,8869705}. Specifically, conventional battery-powered or wired energy provisioning solutions could incur substantial battery replacement and installation expenses, being inconsistent with the vision of green and sustainable communications. Fortunately, wireless energy harvesting (EH) technology \cite{6922609} provides an effective means to alleviate this challenge. By exploiting ambient radio-frequency (RF) signals as a source of wireless energy, low-power IoT devices can replenish their energy supply and support sustained operation without relying on frequent battery replacement or wired connections \cite{6951347}. As one of the most promising technologies for IoT applications, simultaneous wireless information and power transfer (SWIPT) enables the concurrent transmission of information and energy signals, thereby providing wireless information transmission (WIT) services to information receivers (IRs) and wireless power transfer (WPT) to energy receivers (ERs) \cite{6489506,7445146,6957150}. However, in SWIPT systems, ERs may act as potential eavesdroppers, as they are capable of intercepting confidential information intended for IRs without authorization \cite{7462485}. Thus, secure communication in SWIPT systems has become an important research topic. Note that SWIPT systems inherently possess certain physical-layer security (PLS) advantages, since the energy signals can not only supply power to ERs but also serve as artificial noise (AN) to impair their eavesdropping capability \cite{7876873}.
	
	Inevitably, wireless channels introduce severe propagation losses, which can significantly degrade the WPT and WIT performance from the base station (BS) to both the ERs and IRs. To address this challenge, a variety of intelligent antenna technologies have recently been proposed to enhance the transmission capability of SWIPT systems. In \cite{9110849} and \cite{9531372}, the authors employed reconfigurable intelligent surfaces (RISs) to provide additional reflective links for SWIPT systems, thereby facilitating both WIT and WPT. In \cite{10304608,10702481,10525073}, the authors exploited the reflection and transmission capabilities of simultaneously transmitting and reflecting (STAR)-RIS to provide full-space signal enhancement for SWIPT systems. In \cite{11181113,11108293,10840319,11351331,10906511,10753482}, movable antenna (MA) and fluid antenna (FA) technologies were employed to reconfigure the wireless propagation environment, thereby alleviating channel attenuation and improving the WPT and WIT performance of SWIPT. In \cite{11106459,11278178,11206485,11364174}, pinching antenna (PA) technology was exploited by adjusting the antenna position along the waveguide to reduce wireless channel attenuation, while simultaneously enabling pinching beamforming, thereby enhancing the performance of both WPT and WIT. To further leverage the available spatial degrees of freedom, a six-dimensional (6D) MA architecture employing directional antennas has been proposed, in which the position and orientation of the antennas were jointly optimized to improve communication performance \cite{10945745,11142311}. Leveraging the high directional gain of directional antennas and the increased spatial multiplexing gain enabled by 6DMA systems, significant performance improvements can be achieved compared with conventional fixed-antenna systems employing isotropic antennas, through the joint optimization of BS beamforming, position and rotation of antennas \cite{10752873}. It is worth noting that, despite its remarkable performance gains, the distinctive architecture of 6DMA systems also imposes certain limitations on their flexibility.
	
	In this context, a rotatable antenna (RA) architecture with
	lower implementation complexity and greater deployment flexibility has emerged, enabling cost-effective performance enhancement in wireless communications \cite{11301791,zheng2026rotatable,11222668,11251338,wu2026antenna,11427014,11098736}. In RA systems, the orientation/boresight of each directional antenna can be independently adjusted, while the antenna positions remain fixed. Compared to MA/6DMA systems, RA does not require additional physical space for translational motion of individual antennas or antenna arrays. Moreover, RA relies only on local rotation for mechanical or electronic control, which significantly facilitates practical implementation and engineering deployment \cite{11301791,11222668,11251338,wu2026antenna}. Specifically, the authors of \cite{11427014} established an optimization framework for RA systems, and further developed dedicated optimization schemes tailored to the RA architecture. Simulation results demonstrated that RA systems achieved superior performance gains in wireless communications compared with the conventional fixed-antenna and isotropic-antenna schemes. In \cite{11098736}, the authors demonstrated that RA technology can significantly enhance PLS in wireless communication systems. The authors of \cite{11039664} investigated an RA system based on isotropic antennas, where the antenna array was rotated to alter the positions of individual antenna elements, thereby achieving higher spatial multiplexing gains and enhancing the performance of integrated sensing and communication (ISAC) systems.
	
	Despite the growing interest in RA technologies, existing studies on RA systems are still in their early stages. The potential of RA for SWIPT systems has not yet been sufficiently explored. Investigating how RA affects WPT and WIT is therefore of considerable interest. Motivated by these observations, this work focuses on RA-enabled secure SWIPT. Moreover, since PLS is a critical concern in SWIPT systems where energy receivers may act as potential eavesdroppers, we further consider secure communication requirements while exploiting RA to enhance both WIT and WPT. In general, the main contributions of this work are listed below:
	\begin{itemize}
		\item[$\bullet$] We investigate an RA-enabled secure SWIPT system, in which each RA at the BS can be independently rotated to simultaneously provide WIT to multiple IRs and WPT to multiple ERs. Based on this setup, we formulate an optimization problem that maximizes the sum received power at the ERs while satisfying the security requirements, quality-of-service, and power harvesting constraints of the system. Unlike existing works that focus on secrecy rate maximization, the proposed formulation offers a flexible framework to accommodate heterogeneous security requirements, while explicitly addressing the critical WPT performance in SWIPT systems.
		
		\item[$\bullet$] We propose a performance-oriented successive convex approximation (SCA)-based alternating optimization (AO) algorithm to jointly optimize the BS beamforming and RA rotations. Specifically, the original problem is decomposed into two subproblems for BS beamforming and RA rotation optimization. The beamforming subproblem is solved via semidefinite relaxation (SDR) and semidefinite programming (SDP), followed by deterministic rank-one recovery without loss of optimality, while the RA rotation subproblem is tackled using SCA. By alternately solving the two subproblems, the resulting objective sequence is shown to converge. Furthermore, a worst-case robust beamforming extension is developed for norm-bounded channel state information (CSI) errors after fixing the optimized RA orientations.
		
		\item[$\bullet$] We further develop a complexity-oriented differential evolution (DE)-based AO algorithm as an alternative to the performance-oriented SCA-based design. Unlike the SCA method, DE directly searches over the RA rotation angles through mutation, crossover, and selection without requiring channel Jacobians or convex surrogate construction. Although the stochastic nature of DE does not guarantee convergence to a global or local optimum, it offers a simpler implementation and exhibits milder computational growth for RA rotation optimization. Moreover, its population-based search is well suited to the highly non-convex RA rotation problem.
		
		\item[$\bullet$] Numerical results are presented to evaluate the developed algorithms and the proposed RA-enabled secure SWIPT system. Specifically, 1) the performance-oriented SCA-based AO algorithm exhibits fast convergence; 2) the complexity-oriented DE-based AO algorithm achieves competitive performance for small-scale RA arrays, whereas the SCA-based AO algorithm becomes more favorable as the number of RAs increases; 3) the SCA-based and DE-based designs respectively favor solution quality and computational efficiency; and 4) the proposed RA-enabled secure SWIPT system significantly outperforms the isotropic-antenna and fixed-antenna schemes. The effectiveness of the robust beamforming design under imperfect CSI is also verified.
	\end{itemize}
	
	The remainder of this paper is organized as follows. Section \uppercase\expandafter{\romannumeral2} presents the system model and problem formulation for the RA-enabled secure SWIPT system. Sections \uppercase\expandafter{\romannumeral3} and \uppercase\expandafter{\romannumeral4} present the proposed SCA-based and DE-based AO algorithms, respectively. Section \uppercase\expandafter{\romannumeral5} develops the robust beamforming extension under imperfect CSI. Numerical results and performance evaluations are provided in Section \uppercase\expandafter{\romannumeral6}. Finally, Section \uppercase\expandafter{\romannumeral7} concludes the paper.
	
	{\textit{Notations:}} In this paper, italic lowercase letters, bold lowercase letters, and bold uppercase letters denote scalars, vectors, and matrices, respectively. The superscripts $\left(\cdot\right)^T$ and $\left(\cdot\right)^H$ represent the transpose and conjugate transpose operations, respectively. The operators $\left|\cdot\right|$, $\left\|\cdot\right\|$, ${\rm Tr}\left(\cdot\right)$, ${\rm rank}\left(\cdot\right)$, and $\left(\cdot\right) \succeq 0$ denote the absolute value, Euclidean norm, trace, rank, and positive semidefinite constraint, respectively. ${\mathbb R}$ and ${\mathbb C}$ denote the sets of real and complex numbers, respectively. ${\bf I}_N$ denotes the $N \times N$ identity matrix, and ${\bf 0}$ represents an all-zero vector or matrix with appropriate dimensions.
	
	\section{System Model and Problem Formulation}
	\begin{figure}[!t]
		\centering
		\includegraphics[width=3.3in]{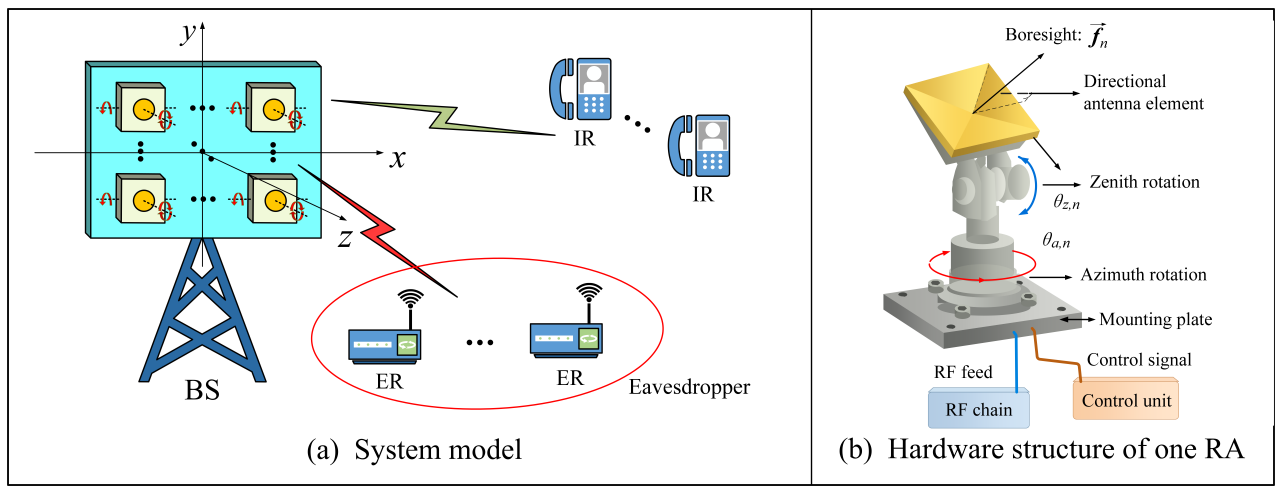}
		\caption{The proposed RA-enabled secure SWIPT system.}
		\label{system_model}
	\end{figure}
	Fig. \ref{system_model} depicts an RA-enabled secure SWIPT system, in which a BS employs a uniform planar array (UPA) formed by $N$ directional RAs to concurrently provide wireless information transmission for $K_I$ single-antenna IRs and wireless power transfer for $K_E$ single-antenna ERs. For notational convenience, we set $n \in {\cal N} \buildrel \Delta \over = \left\{1,...,N\right\}$, $i \in {\cal K_I} \buildrel \Delta \over = \left\{1,...,K_I\right\}$, and $j \in {\cal K_E} \buildrel \Delta \over = \left\{1,...,K_E\right\}${\footnote{The IRs are legitimate receivers authorized to decode the confidential information transmitted by the BS, whereas the ERs are registered EH devices authorized for energy harvesting. Hence, the ERs are modeled as legitimate-but-untrusted receivers that may eavesdrop on the confidential signals intended for IRs \cite{7462485,10525073,11108293}. As registered network devices, the ERs can participate in channel training or limited-feedback procedures, making their CSI more readily obtainable than that of unknown external eavesdroppers \cite{11134688}.}}. 
	We assume that the UPA lies in the plane formed by the $x$- and $y$-axes in the 3D coordinate system. Moreover, we have $N = N_x \times N_y$, where $N_x$ and $N_y$ represent the numbers of the RAs along each row and column of the UPA, respectively. To alleviate mutual coupling between the RAs and ensure sufficient space for their rotation, the center-to-center spacing between adjacent RAs is set to $\Delta$. Accordingly, the aperture of the UPA measured between the centers of the outermost RAs is $(N_x-1)\Delta \times (N_y-1)\Delta$.
	Each RA has a physical size of $\sqrt{A}\times\sqrt{A}$, where $\sqrt{A}\leq\Delta$~\cite{zheng2026rotatable}. The $n$-th RA is associated with a pair of centered array indices $(n_x,n_y)$ and has the 3D coordinate ${{\bf z}_n}\buildrel\Delta\over={{\bf z}_{n_x,n_y}}=\left[n_x\Delta,n_y\Delta+H_{\rm BS},0\right]^T$, where
	$n_x\in\left\{m-\frac{N_x-1}{2}\mid m=0,\ldots,N_x-1\right\}$ and
	$n_y\in\left\{l-\frac{N_y-1}{2}\mid l=0,\ldots,N_y-1\right\}$. Here, $H_{\rm BS}$ denotes the height of the UPA center above the ground. Both the IRs and ERs are considered ground-based devices. Accordingly, the 3D coordinates of the $i$-th IR and $j$-th ER can be expressed as ${\bf u}_i \buildrel \Delta \over = \left[u_{x,i}, 0, u_{z,i}\right]^T$ and ${\bf u}_j \buildrel \Delta \over = \left[u_{x,j}, 0, u_{z,j}\right]^T$, respectively. 
	To account for the physical swept volume of each independently rotatable element, let $r_{\rm RA}$ denote the radius of a minimum bounding sphere containing one RA over its entire feasible rotation range. The antenna spacing is selected offline to satisfy $\Delta  \ge 2{r_{{\rm{RA}}}} + {d_{{\rm{safe}}}}$, where $d_{{\rm{safe}}} > 0$ is a prescribed safety margin\footnote{Since the RA positions and array spacing are fixed, the above geometric condition can be enforced offline during array design and remains satisfied for all feasible RA rotations, without requiring additional angle-dependent constraints. Consequently, the swept volumes of adjacent RAs do not overlap, which prevents physical collision during antenna rotation and provides sufficient spacing to mitigate inter-antenna blockage and coupling \cite{11222668,zheng2026rotatable,11251338}.}.
	
	
	\subsection{Channel Model}
	In this work, each RA within the UPA can be independently controlled to perform antenna rotation. The initial pointing vector of each RA is assumed to be aligned with the direction of the $z$-axis. For convenience, the pointing vector of the $n$-th RA is represented by 
	\begin{flalign}
		\vec{\bf f}_n = {\left[ {{f_{x,n}},{f_{y,n}},{f_{z,n}}} \right]^T} \in {{\mathbb R}^{3 \times 1}},
	\end{flalign}
	where $\big\| {{\vec{\bf{f}}_n}} \big\| = 1$ indicates that $\vec{\bf f}_n$ is a normalized direction vector. In addition, ${f_{x,n}} = \sin {\theta _{z,n}}\cos {\theta _{a,n}}$, ${f_{y,n}} = \sin {\theta _{z,n}}\sin {\theta _{a,n}}$, and ${f_{z,n}} = \cos {\theta _{z,n}}$ represent the components of ${\bf f}_n$ along the $x$-, $y$-, and $z$-axes, respectively. Here, $\theta_{z,n} \in \left[0, \theta_{\max}\right]$ and $\theta_{a,n} \in \left[0, 2 \pi\right)$ stand for the zenith and azimuth angles of the $n$-th RA, respectively, while $\theta_{\max} \in \left[0, \frac{\pi}{2}\right]$ denotes the maximum zenith angle supported by each RA. 
	
	In the considered system, the channel gains of the RA-to-IR and RA-to-ER links are determined by the signal transmission/reception angles of the RAs. Therefore, each RA is characterized by the generic directional gain pattern \cite{11427014} that describes its angle-dependent channel gain, which is given by
	\begin{flalign}
		{G_e}\left( {\varepsilon ,\varphi } \right) = \left\{ {\begin{array}{*{20}{c}}
				{{G_0}{{\cos }^{2p}}\left( \varepsilon  \right),}&{\varepsilon  \in \left[ {0,\frac{\pi }{2}} \right],\varphi  \in \left[ {0,2\pi } \right)}\\
				0,&{{\rm{otherwise,}}}
		\end{array}} \right. \label{Ge}
	\end{flalign}
	where $(\varepsilon,\varphi)$ denotes the angular offset between the RA boresight and the direction from the RA to the receiving device. For the directional-antenna model, $p>0$ characterizes the directivity and governs the main-lobe beamwidth. The boresight gain is $G_0=2(2p+1)$, which normalizes the radiated power over the front hemisphere. Due to rotational symmetry about the boresight, the gain depends only on $\varepsilon${\footnote{The adopted radiation pattern is a tractable main-lobe approximation that captures the dominant orientation-dependent gain of each RA, while side- and back-lobe responses are neglected. In practice, a measured or calibrated radiation pattern can be incorporated by replacing $G(\varepsilon)$ with the corresponding pattern function and adapting the associated optimization expressions accordingly.}}. For the ideal isotropic-antenna benchmark, we separately set $G_e^{\rm iso}(\varepsilon,\varphi)=1$ over the full sphere, $G_0=1$, and the polarization-overlap factor to one. Hence, this benchmark is not obtained by directly substituting $p=0$ into the one-sided pattern in \eqref{Ge}.
	
	In this work, we consider a finite-scattering propagation environment containing one LoS path and $Q$ single-bounce NLoS paths. Let ${\cal Q} \buildrel \Delta \over = \left\{ {1, \ldots ,Q} \right\}$ denote the scatterer index set, and let ${\bf s}_q \in {\mathbb{R}}^{3 \times 1}$ denote the position of scatterer $q$.
		The unit direction vectors from the $n$-th RA to the $i$-th IR and the $j$-th ER are respectively given by ${\vec {\bf{q}} _{n,i}} \buildrel \Delta \over = \frac{{{{\bf{u}}_i} - {{\bf{z}}_n}}}{{\left\| {{{\bf{u}}_i} - {{\bf{z}}_n}} \right\|}}$
		and
		${\vec {\bf{q}} _{n,j}} \buildrel \Delta \over = \frac{{{{\bf{u}}_j} - {{\bf{z}}_n}}}{{\left\| {{{\bf{u}}_j} - {{\bf{z}}_n}} \right\|}}$, 
		which are normalized direction vectors. For the $q$-th NLoS path, the corresponding departure direction from the $n$-th RA to scatterer $q$ is ${\vec {\bf{q}} _{n,q}} \buildrel \Delta \over = \frac{{{{\bf{s}}_q} - {{\bf{z}}_n}}}{{\left\| {{{\bf{s}}_q} - {{\bf{z}}_n}} \right\|}}$. The corresponding distances are ${d_{n,i}} = \left\| {{{\bf{z}}_n} - {{\bf{u}}_i}} \right\|$, ${d_{n,j}} = \left\| {{{\bf{z}}_n} - {{\bf{u}}_j}} \right\|$, ${d_{n,q}} = \left\| {{{\bf{z}}_n} - {{\bf{s}}_q}} \right\|$, ${d_{q,i}} = \left\| {{{\bf{s}}_q} - {{\bf{u}}_i}} \right\|$, and ${d_{q,j}} = \left\| {{{\bf{s}}_q} - {{\bf{u}}_j}} \right\|$.
		
		Following the RA channel modeling framework in \cite{zheng2026rotatable,11427014}, the orientation-independent complex coefficient of the LoS path from RA $n$ to IR $i$ is defined as
		\begin{flalign}
			\alpha _{n,i}^{\rm L}
			=
			\sqrt{{A_0}{G_0}d_{n,i}^{-{\alpha_i}}}
			\exp\left(-j\frac{2\pi}{\lambda}d_{n,i}\right).
		\end{flalign}
		The coefficient $\alpha_{n,j}^{\rm L}$ for ER $j$ is defined analogously. The complex coefficient of the $q$-th single-bounce NLoS path from RA $n$ to IR $i$ is modeled as
		\begin{flalign}
			\alpha _{n,i,q}^{\rm N}
			={}&
			\sqrt{{A_0}{G_0}{\zeta_{i,q}}
				d_{n,q}^{-{\alpha_q}}
				d_{q,i}^{-{\alpha_{q,i}}}}
			\times \notag\\
			&\exp\left[
			-j\frac{2\pi}{\lambda}
			\left(d_{n,q}+d_{q,i}\right)
			+j\phi_{i,q}
			\right],
		\end{flalign}
		where $\zeta_{i,q}$ and $\phi_{i,q}$ denote the power reflection coefficient and phase shift, respectively. The coefficients $\alpha_{n,j}^{\rm L}$ and $\alpha_{n,j,q}^{\rm N}$ are defined analogously for the $j$-th ER. Here, $A_0$ denotes the channel power gain at the reference distance, $\lambda$ is the carrier wavelength, and $\alpha_i$, $\alpha_q$, and $\alpha_{q,i}$ are the path-loss exponents of the corresponding propagation segments; the ER-side exponents are defined analogously.
		
		Antenna rotation may additionally cause polarization mismatch. We assume that the RAs and the receiver antennas are linearly polarized. Following the polarization-aware channel model in Appendix E of \cite{11427014}, the main polarization direction of RA $n$ after rotation is expressed as $\mathbf p_n(\boldsymbol\theta_n) =\left[\cos\theta_{z,n}\cos\theta_{a,n},\, \cos\theta_{z,n}\sin\theta_{a,n},\,-\sin\theta_{z,n}\right]^T$, where $\boldsymbol\theta_n\triangleq[\theta_{z,n},\theta_{a,n}]^T$ and $\mathbf p_n^T(\boldsymbol\theta_n)\vec{\mathbf f}_n=0$. For a generic receiver $r$, where $r=i$ for an IR and $r=j$ for an ER, the projection of $\mathbf p_n(\boldsymbol\theta_n)$ onto the tangent plane orthogonal to the LoS propagation direction is given by $\mathbf p_{n,r}^{\mathrm{r,L}}(\boldsymbol\theta_n) =\mathbf p_n(\boldsymbol\theta_n) -\left(\mathbf p_n^T(\boldsymbol\theta_n)\vec{\mathbf q}_{n,r}\right) \vec{\mathbf q}_{n,r}$. Similarly, the projection associated with the $q$-th NLoS path is $\mathbf p_{n,r,q}^{\mathrm{r,N}}(\boldsymbol\theta_n) =\mathbf p_n(\boldsymbol\theta_n) -\left(\mathbf p_n^T(\boldsymbol\theta_n)\vec{\mathbf q}_{n,q}\right) \vec{\mathbf q}_{n,q}$. Let $\mathbf p_{n,r}^{\mathrm{t,L}}$ and $\mathbf p_{n,r,q}^{\mathrm{t,N}}$ denote the effective electric-field vectors of the LoS and reflected NLoS waves, respectively. By channel reciprocity, the polarization-overlap coefficients for the downlink RA-to-receiver channels can be written as $\varrho_{n,r}^{\mathrm L}(\boldsymbol\theta_n) =\left(\mathbf p_{n,r}^{\mathrm{t,L}}\right)^T \mathbf p_{n,r}^{\mathrm{r,L}}(\boldsymbol\theta_n)$ and $\varrho_{n,r,q}^{\mathrm N}(\boldsymbol\theta_n) =\left(\mathbf p_{n,r,q}^{\mathrm{t,N}}\right)^T \mathbf p_{n,r,q}^{\mathrm{r,N}}(\boldsymbol\theta_n)$, where the reflection-induced polarization transformation is incorporated into $\mathbf p_{n,r,q}^{\mathrm{t,N}}$. Under this reciprocal representation, the path electric-field vectors are fixed for a given channel realization. Accordingly, the channels from the $n$-th RA to the $i$-th IR and the $j$-th ER are respectively written as
		\begin{flalign}
			h_{n,i}(\boldsymbol\theta_n) ={}
			&\varrho_{n,i}^{\mathrm L}(\boldsymbol\theta_n)\alpha_{n,i}^{\mathrm L} \left[\vec{\mathbf f}_n^T(\boldsymbol\theta_n)\vec{\mathbf q}_{n,i}\right]_{+}^{p} \notag\\
			&+\sum\nolimits_{q=1}^{Q}\varrho_{n,i,q}^{\mathrm N}(\boldsymbol\theta_n)\alpha_{n,i,q}^{\mathrm N} \left[\vec{\mathbf f}_n^T(\boldsymbol\theta_n)\vec{\mathbf q}_{n,q}\right]_{+}^{p}, \label{hni}\\ g_{n,j}(\boldsymbol\theta_n) ={}
			&\varrho_{n,j}^{\mathrm L}(\boldsymbol\theta_n)\alpha_{n,j}^{\mathrm L} \left[\vec{\mathbf f}_n^T(\boldsymbol\theta_n)\vec{\mathbf q}_{n,j}\right]_{+}^{p} \notag\\
			&+\sum\nolimits_{q=1}^{Q}\varrho_{n,j,q}^{\mathrm N}(\boldsymbol\theta_n)\alpha_{n,j,q}^{\mathrm N}
			\left[\vec{\mathbf f}_n^T(\boldsymbol\theta_n)\vec{\mathbf q}_{n,q}\right]_{+}^{p}.
			\label{gnj}
		\end{flalign}
		where $[x]_{+}\triangleq\max\{x,0\}$. Setting $Q=0$ yields the pure LoS model, whereas setting all polarization-overlap coefficients to one yields the polarization-matched model \cite{11427014}.
		
		Accordingly, the combined channel vectors from the BS to the $i$-th IR and the $j$-th ER are respectively written as
		\begin{flalign}
			&\mathbf h_i(\boldsymbol\theta)=[h_{1,i}(\boldsymbol\theta_1),\ldots,h_{N,i}(\boldsymbol\theta_N)]^T,\label{hi}\\
			&\mathbf g_j(\boldsymbol\theta)=[g_{1,j}(\boldsymbol\theta_1),\ldots,g_{N,j}(\boldsymbol\theta_N)]^T,\label{gj}
		\end{flalign}
		where $\boldsymbol\theta=[\theta_{z,1},\theta_{a,1},\ldots,\theta_{z,N},\theta_{a,N}]^T$, and $\mathbf F(\boldsymbol\theta)=[\vec{\mathbf f}_1(\boldsymbol\theta_1),\ldots,\vec{\mathbf f}_N(\boldsymbol\theta_N)]\in\mathbb R^{3\times N}$.
		
		\textit{Remark 1:} The RA orientations are updated once at the beginning of each quasi-static channel block and remain fixed during the subsequent transmission. The proposed algorithms are executed numerically at the BS and do not require physical rotation at every iteration. Hence, mechanically driven RAs are more suitable for static or low-mobility scenarios, whereas electronically driven RAs can support faster reconfiguration. The block-wise design is applicable when the channel-estimation, optimization, and RA reconfiguration time is shorter than the channel coherence time \cite{zheng2026rotatable,11222668}. For high-mobility users, channel aging and reconfiguration delay may cause orientation-mismatch losses. Such degradation depends on the user velocity, channel coherence time, and RA hardware characteristics. The actuation energy of mechanical RAs is also hardware dependent and is not included in the RF transmit-power budget. A joint treatment of mobility, reconfiguration delay, and actuation energy is left for future work.

	\subsection{Signal Model}
	In the considered system, the IRs are assumed to be legitimate devices that receive and decode confidential information signals transmitted from the BS. In contrast, the ERs are considered potential eavesdroppers, as they may attempt to intercept the information intended for the IRs. To enhance communication security, AN is intentionally embedded in the transmitted signal of the BS. This AN not only interferes with the ability of ERs to decode the confidential information, but also contributes to their energy harvesting. Therefore, the composite signal transmitted by the BS can be expressed as 
	\begin{flalign}
		{\bf{x}} = \sum\nolimits_{i = 1}^{{K_I}} {{{\bf{w}}_i}{s_i}}  + {\bf{v}},
	\end{flalign}
	where ${\bf w}_i \in {\mathbb C}^{N \times 1}$ denotes the transmit beamforming vector associated with the $i$-th IR, and ${s_i} \in {\mathbb C}$ is the corresponding information symbol sent by the BS. The information symbols are mutually independent and normalized such that $\mathbb E\{|s_i|^2\}=1$, $\forall i$. In addition, ${\bf v} \in {\mathbb C}^{N \times 1}$ stands for the AN injected at the BS, which is modeled as a circularly symmetric complex Gaussian random vector, i.e., ${\bf v} \sim {\cal CN}\left({\bf 0},{\bf V}\right)$ with ${\bf V} \succeq {\bf 0}$. Accordingly, the composite signal received at the $i$-th IR can be written as
	\begin{flalign}
		{y_i} = {\bf{h}}_i^H(\boldsymbol\theta){\bf{x}} + {n_i},
	\end{flalign}
	where $n_i\sim{\cal CN}(0,\sigma_i^2)$ denotes the additive white Gaussian noise (AWGN) at the $i$-th IR. Hence, the signal-to-interference-plus-noise ratio (SINR) of the signal received by the $i$-th IR can be given by
	\begin{flalign}
		{\gamma _i^{\rm IR}} = \frac{{{{\left| {{\bf{h}}_i^H(\boldsymbol\theta){{\bf{w}}_i}} \right|}^2}}}{{\sum\nolimits_{k \ne i}^{{K_I}} {{{\left| {{\bf{h}}_i^H(\boldsymbol\theta){{\bf{w}}_k}} \right|}^2}}  + {\bf{h}}_i^H(\boldsymbol\theta){\bf{V}}{{\bf{h}}_i(\boldsymbol\theta)} + \sigma _i^2}},
	\end{flalign}
	where $\sigma _i^2$ is the power of the AWGN at the $i$-th IR. Since the primary function of the ERs is energy harvesting rather than information decoding, they can collect energy from all components of the received signal. Furthermore, the power of the additive noise is typically negligible compared to that of the signal received at the ERs. Accordingly, its impact on the harvested energy is ignored in this work. Thus, the received signal power at the $j$-th ER can be given by
	\begin{flalign}
		{P_j} = {\bf{g}}_j^H(\boldsymbol\theta)\left( {\sum\nolimits_{i = 1}^{{K_I}} {{{\bf{w}}_i}{\bf{w}}_i^H}  + {\bf{V}}} \right){{\bf{g}}_j}(\boldsymbol\theta).
	\end{flalign}
	
	The ERs employ rectification circuits to transform the received RF power into usable direct-current (DC) energy for local operation. Nevertheless, this RF-to-DC conversion is inevitably accompanied by energy loss. To characterize the practical nonlinear behavior of the energy harvesting process, we adopt the nonlinear EH model in \cite{11206485,7264986}. Accordingly, the harvested power at the $j$-th ER is given by 
	\begin{flalign}
		{\Phi _j} = \frac{{{\Psi _j} - {Z_j}{\Omega _j}}}{{1 - {\Omega _j}}},~~~~~~~~~~~~~\\
		{\Psi _j} = \frac{{{Z_j}}}{{1 + {e^{ - {a_j}\left( {{P_j} - {b_j}} \right)}}}},~{\Omega _j} = \frac{1}{{1 + {e^{{a_j}{b_j}}}}},
	\end{flalign}
	where $a_j$, $b_j$, and $Z_j$ are the hardware parameters of the energy harvester at the $j$-th ER. Meanwhile, the ERs are considered potential eavesdroppers that may attempt to intercept the confidential signals transmitted from the BS to the IRs. When the ERs attempt to eavesdrop on the confidential signals transmitted to the IRs, the SINR at the $j$-th ER for the signal intended for the $i$-th IR can be given by
	\begin{flalign}
		{\gamma _{i,j}^{\rm ER}} = \frac{{{{\left| {{\bf{g}}_j^H(\boldsymbol\theta){{\bf{w}}_i}} \right|}^2}}}{{\sum\nolimits_{k \ne i}^{{K_I}} {{{\left| {{\bf{g}}_j^H(\boldsymbol\theta){{\bf{w}}_k}} \right|}^2}}  + {\bf{g}}_j^H(\boldsymbol\theta){\bf{V}}{{\bf{g}}_j}(\boldsymbol\theta) + \sigma _j^2}}.
	\end{flalign}
	where $\sigma_j^2$ denotes the AWGN power at the $j$-th ER.
	
	\subsection{Problem Formulation}
	We aim to maximize the sum received power at the ERs through the joint design of the RA rotation angles and the transmit beamforming at the BS. The design is subject to the minimum SINR requirements of the IRs, the minimum harvested-energy requirements of the ERs, and the upper bounds on the eavesdropping SINR at the ERs for secure transmission. The received power $P_j$ is adopted as the WPT performance metric, whereas the practical nonlinear RF-to-DC conversion is explicitly enforced through the harvested-power constraint $\Phi_j\geq P_{\min}$. Define ${\bf w}=\{{\bf w}_1,\ldots,{\bf w}_{K_I}\}$. Then, the corresponding sum received power maximization problem is formulated as
	\begin{subequations}\label{p1}
		\begin{align}
			\mathop {\max }\limits_{{\boldsymbol\theta},{\bf{w}},{{\bf V} \succeq {\bf 0}}}~ &\sum\nolimits_{j = 1}^{{K_E}} P_j  \label{p1-a}\\
			{\rm s.t.~~~}~&\gamma _i^{{\rm{IR}}} \ge \gamma _{\min }^{\rm IR},\forall i, \label{p1-b}\\
			&{\Phi _j} \ge {P_{\min }},\forall j, \label{p1-c}\\
			&\gamma _{i,j}^{{\rm{ER}}} \le {\gamma _{\max }^{\rm ER}},\forall i,j, \label{p1-d}\\
			&\sum\nolimits_{i = 1}^{{K_I}} {{{\left\| {{{\bf{w}}_i}} \right\|}^2}}  + {\rm{Tr}}\left( {\bf{V}} \right) \le {P_0}, \label{p1-e}\\
			&0\leq\theta_{z,n}\leq\theta_{\max},\quad
				0\leq\theta_{a,n}<2\pi,\quad\forall n, \label{p1-f}
		\end{align}
	\end{subequations}
	where constraints (\ref{p1-b}) and (\ref{p1-c}) require that the received SINR at the IRs and the harvested DC power at the ERs meet the minimum requirements $\gamma _{\min}^{\rm IR}$ and $P_{\min}$, respectively. Constraint (\ref{p1-d}) stipulates that the eavesdropping SINR at the ERs must not exceed $\gamma _{\max}^{\rm ER}$ in order to ensure secure communication. (\ref{p1-e}) is the transmit power constraint of the BS. The remaining constraints have been described previously and are therefore not discussed further here.
	
	\textit{Remark 2:} Conventional electronic beamforming with fixed directional antennas is a special case of problem~\eqref{p1}, obtained by fixing $\vec{\mathbf f}_n=[0,0,1]^T$ for all $n$. Hence, RA rotation complements electronic beamforming by providing an additional orientation-domain DoF through element-level directional gain reconfiguration. In Section VI, a fixed-orientation directional-antenna scheme with the same number of antennas, array aperture, and transmit-power budget is adopted as a benchmark.
	
	It can be readily seen that the variable ${\bm \theta}$ is strongly coupled with ${\bf w}$ and $\bf V$, which renders both the objective function and constraints (\ref{p1-b})-(\ref{p1-d}) highly intractable and non-convex.
	
	\section{The Proposed SCA-based AO Algorithm}
	To tackle these difficulties, we decompose problem~\eqref{p1} into beamforming and RA-rotation subproblems and solve them alternately. In each AO iteration, the beamforming variables are updated for the current RA orientations, followed by one SCA update for RA rotation optimization with the beamforming variables fixed. The two subproblems are then solved iteratively until convergence.
	
	\subsection{Beamforming Optimization for RAs}
	To facilitate the subsequent solution process, we first reformulate problem (\ref{p1}). Let ${\bf W}_i =  {\bf w}_i {\bf w}_i^H$, which must satisfy ${\bf W}_i \succeq {\bf 0}$ and ${\rm rank}{\left({\bf W}_i\right)} = 1$. Accordingly, let ${\bf W} = \left\{{\bf W}_1,...{\bf W}_{K_I}\right\}$, problem (\ref{p1}) can be recast as
	\begin{subequations}\label{p2}
		\begin{align}
			&\mathop{\max}\limits_{{\left\{{\bf W}_i\succeq {\bf 0}\right\}_{i=1}^{K_I}},\hfill\atop
				\scriptstyle {\bm \theta},{\bf V} \succeq {\bf 0}} \quad  \sum\nolimits_{j = 1}^{{K_E}} P_j \label{p2-a} \\
			\text{s.t.~~~}  &\frac{{{\rm{Tr}}\left( {{{\bf{H}}_i}{{\bf{W}}_i}} \right)}}{{\sum\nolimits_{k \ne i}^{{K_I}} {{\rm{Tr}}\left( {{{\bf{H}}_i}{{\bf{W}}_k}} \right)}  + {\rm{Tr}}\left( {{{\bf{H}}_i}{\bf{V}}} \right) + \sigma _i^2}} \ge \gamma _{\min }^{\rm IR},\forall i, \label{p2-b}\\
			&{\rm Tr}\left(\sum\nolimits_{i = 1}^{{K_I}} {{\bf{G}}_j{{\bf{W}}_i}}  + {\bf{G}}_j{\bf V}\right) \ge B_j,\forall j, \label{p2-c}\\
			&\frac{{{\rm{Tr}}\left( {{{\bf{G}}_j}{{\bf{W}}_i}} \right)}}{{\sum\nolimits_{k \ne i}^{{K_I}} {{\rm{Tr}}\left( {{{\bf{G}}_j}{{\bf{W}}_k}} \right)}  + {\rm{Tr}}\left( {{{\bf{G}}_j}{\bf{V}}} \right) + \sigma _j^2}} \le {\gamma _{\max }^{\rm ER}},\forall i,j, \label{p2-d}\\
			&\sum\nolimits_{i = 1}^{{K_I}} {{\rm{Tr}}\left( {{{\bf{W}}_i}} \right)}  + {\rm{Tr}}\left( {\bf{V}} \right) \le {P_0}, \label{p2-e}\\
			&{\rm rank}\left({\bf W}_i\right) = 1, \forall i, \label{p2-f}\\
			&\left(\rm \ref{p1-f}\right),
		\end{align}
	\end{subequations}
	where $B_j = {b_j} - \frac{{\ln \left( {\frac{{{Z_j}}}{{{P_{\min }}\left( {1 - {\Omega _j}} \right) + {Z_j}{\Omega _j}}} - 1} \right)}}{{{a_j}}}$ is the minimum received RF power required to satisfy the nonlinear EH constraint, assuming that the prescribed $P_{\min}$ lies within the feasible output range of the rectifier. Moreover, ${\bf H}_i(\bm \theta) = {\bf h}_i(\bm \theta) {\bf h}_i^H(\bm \theta)$ and ${\bf G}_j(\bm \theta) = {\bf g}_j(\bm \theta) {\bf g}_j^H(\bm \theta)$. The reformulated problem (\ref{p2}) remains non-convex because ${\bm \theta}$ is coupled with ${\bf W}$ and ${\bf V}$ through the channel matrices ${\bf H}_i(\bm\theta)$ and ${\bf G}_j(\bm\theta)$. For a given ${\bm \theta}$, the channel matrices are fixed, and the SINR constraints can be equivalently rewritten. Thus, problem (\ref{p2}) can be recast as
	\begin{subequations}\label{p3} 
		\begin{align}
			&\mathop{\max}\limits_{{\left\{{\bf W}_i\succeq {\bf 0}\right\}_{i=1}^{K_I}},\hfill\atop
				\scriptstyle {\bf V} \succeq {\bf 0}}  \sum\limits_{j = 1}^{K_E}{\rm Tr}\left(\sum\nolimits_{i = 1}^{{K_I}} {{\bf{G}}_j(\bm \theta){{\bf{W}}_i}}  + {\bf{G}}_j(\bm \theta){\bf V}\right) \label{p3-a} \\
			&\text{s.t.~} {\rm{Tr}}\left( {{{\bf{H}}_i(\bm \theta)}{{\bf{W}}_i}} \right) \ge \gamma _{\min }^{{\rm{IR}}}\sum\nolimits_{k \ne i}^{{K_I}} {{\rm{Tr}}\left( {{{\bf{H}}_i(\bm \theta)}{{\bf{W}}_k}} \right)} \notag\\
			& ~~~~~~~~+ \gamma _{\min }^{{\rm{IR}}}{\rm{Tr}}\left( {{{\bf{H}}_i(\bm \theta)}{\bf{V}}} \right) + \gamma _{\min }^{{\rm{IR}}}\sigma _i^2,\forall i, \label{p3-b}\\
			&~~~~{\rm{Tr}}\left( {{{\bf{G}}_j(\bm \theta)}{{\bf{W}}_i}} \right) \le {\gamma _{\max }^{{\rm{ER}}}}\sum\nolimits_{k \ne i}^{{K_I}} {{\rm{Tr}}\left( {{{\bf{G}}_j(\bm \theta)}{{\bf{W}}_k}} \right)} \notag\\
			& ~~~~~~~~+ {\gamma _{\max }^{{\rm{ER}}}}{\rm{Tr}}\left( {{{\bf{G}}_j(\bm \theta)}{\bf{V}}} \right) + {\gamma _{\max }^{{\rm{ER}}}}\sigma _j^2,\forall i,j, \label{p3-d}\\
			&~~~~{\eqref{p2-c}},{\eqref{p2-e}},{\eqref{p2-f}}.
		\end{align}
	\end{subequations}
	
	However, problem (\ref{p3}) remains hard to address since the non-convex rank-one limitation in (\ref{p2-f}). For the aforementioned challenge, we first apply SDR to disregard constraint
	(\ref{p2-f}). As a result, the SDR problem (\ref{p3}) becomes a convex problem with respect to the variables ${\bf W}$ and ${\bf V}$, which can be directly solved using various existing convex optimization methods, such as SDP \cite{boyd2004convex}.
	Subsequently, the rank-one constraint that is relaxed needs to be restored. Based on \textit{Theorem~1}, an optimal solution $\{\mathbf W_i^\star\}$ satisfying the rank-one constraints can always be constructed from the optimal SDR solution without loss of optimality. Finally, the corresponding beamforming vector $\mathbf w_i^\star$ can be recovered according to \eqref{rank_one_construction}.
	
	\smallskip
	{\textit{Theorem 1:}} If the SDR of problem~(\ref{p3}) is feasible, there always exists an optimal solution
	$\{\mathbf W_i^\star\}_{i=1}^{K_I}$ and $\mathbf V^\star$ to the SDR
	satisfying ${\rm rank}(\mathbf W_i^\star)=1$, $\forall i$.
	
	\begin{proof}
		The proof details can be found in Appendix A.
	\end{proof}

	\subsection{Rotation Optimization for RAs}
	Given ${\bf w}$ and ${\bf V}$ from the above subsection, we optimize the RA rotation angles. For fixed ${\bf w}$ and ${\bf V}$, problem (\ref{p1}) can be re-expressed as
		\begin{subequations}\label{p4}
			\begin{align}
				\mathop {\max }\limits_{\boldsymbol\theta}\quad
				&\sum\nolimits_{j = 1}^{{K_E}} { {{{\bf{g}}_j^H(\boldsymbol\theta)}\left( {\sum\nolimits_{i = 1}^{{K_I}} {{{\bf{w}}_i}{{\bf{w}}_i^H}}  + {\bf{V}}} \right){{\bf{g}}_j(\boldsymbol\theta)}}}  \label{p4-a}\\
				{\rm s.t.}\quad
				&\left(\rm \ref{p1-b}\right), \left(\rm \ref{p1-c}\right), \left(\rm \ref{p1-d}\right), \label{p4-b}\\
				&0\leq\theta_{z,n}\leq\theta_{\max},\quad
					0\leq\theta_{a,n}<2\pi,\quad\forall n. \label{p4-c}
			\end{align}
		\end{subequations}
		Problem \eqref{p4} remains non-convex because the channel vectors are nonlinear functions of the RA rotation angles.
		
		{\textit{Remark 3:}} For the SCA-based design, the positive-part operator in \eqref{hni} and \eqref{gnj} is retained because a propagation direction may move outside the front hemisphere as the RA rotates. Therefore, the local quadratic approximations are constructed directly from the original channel expressions within a prescribed trust region, where the curvature parameters are introduced to control the local approximation behavior.
		
		To obtain the channel Jacobians, define
		$\mathbf b_{n,z}\triangleq\partial\vec{\mathbf f}_n/\partial\theta_{z,n}=\mathbf p_n$,
		$\mathbf b_{n,a}\triangleq\partial\vec{\mathbf f}_n/\partial\theta_{a,n}
		=[-\sin\theta_{z,n}\sin\theta_{a,n},\,\sin\theta_{z,n}\cos\theta_{a,n},\,0]^T$,
		$\mathbf d_{n,z}\triangleq\partial\mathbf p_n/\partial\theta_{z,n}
		=-\vec{\mathbf f}_n$, and
		$\mathbf d_{n,a}\triangleq\partial\mathbf p_n/\partial\theta_{a,n}
		=[-\cos\theta_{z,n}\sin\theta_{a,n},\,\cos\theta_{z,n}\cos\theta_{a,n},\,0]^T$.
		For $s\in\{z,a\}$, the derivatives of the polarization-overlap coefficients are
		$\dot\varrho_{n,r,s}^{\mathrm L}
		=(\mathbf p_{n,r}^{\mathrm{t,L}})^T
		(\mathbf I_3-\vec{\mathbf q}_{n,r}\vec{\mathbf q}_{n,r}^T)\mathbf d_{n,s}$
		and
		$\dot\varrho_{n,r,q,s}^{\mathrm N}
		=(\mathbf p_{n,r,q}^{\mathrm{t,N}})^T
		(\mathbf I_3-\vec{\mathbf q}_{n,q}\vec{\mathbf q}_{n,q}^T)\mathbf d_{n,s}$.
		At the $t$-th AO iteration, the derivative of the channel coefficient from RA $n$ to a generic receiver $r$ is given by
		\begin{align}
				&\xi_{n,r,s}^{(t)}
				={}\alpha_{n,r}^{\mathrm L}\left[
				\dot\varrho_{n,r,s}^{\mathrm L}[a_{n,r}]_{+}^{p}
				+p\varrho_{n,r}^{\mathrm L}[a_{n,r}]_{+}^{p-1}
				\vec{\mathbf q}_{n,r}^{T}\mathbf b_{n,s}\right]+\nonumber\\
				&\sum\limits_{q=1}^{Q}\alpha_{n,r,q}^{\mathrm N}\!\left[
				\dot\varrho_{n,r,q,s}^{\mathrm N}[a_{n,q}]_{+}^{p}
				\!+\!p\varrho_{n,r,q}^{\mathrm N}[a_{n,q}]_{+}^{p-1}
				\vec{\mathbf q}_{n,q}^{T}\mathbf b_{n,s}\right],
			\label{31}
		\end{align}
		where $a_{n,r}=\vec{\mathbf f}_n^T\vec{\mathbf q}_{n,r}$ and
		$a_{n,q}=\vec{\mathbf f}_n^T\vec{\mathbf q}_{n,q}$, and all quantities on the right-hand side of \eqref{31} are evaluated at $\boldsymbol\theta^{(t)}$.
		Let $\bar{\mathbf e}_n\in\mathbb R^N$ denote the $n$-th canonical basis vector.
		The corresponding Jacobian block is
		$\mathbf J_{r,n}^{(t)}
		=\bar{\mathbf e}_n[\xi_{n,r,z}^{(t)},\xi_{n,r,a}^{(t)}]$,
		and
		$\mathbf J_r^{(t)}
		=[\mathbf J_{r,1}^{(t)},\ldots,\mathbf J_{r,N}^{(t)}]$.
		
		For compactness, let $\mathbf c_r(\boldsymbol\theta)$ denote either
		$\mathbf h_i(\boldsymbol\theta)$ or $\mathbf g_j(\boldsymbol\theta)$, and define the real-valued received-power function
		$u_{r,\mathbf A}(\boldsymbol\theta)
		\triangleq
		\mathbf c_r^H(\boldsymbol\theta)\mathbf A\mathbf c_r(\boldsymbol\theta)$
		for any fixed Hermitian positive semidefinite matrix $\mathbf A$.
		Its gradient at $\boldsymbol\theta^{(t)}$ is
		$\nabla u_{r,\mathbf A}(\boldsymbol\theta^{(t)})
		=
		2\mathrm{Re}\{(\mathbf J_r^{(t)})^H
		\mathbf A\mathbf c_r(\boldsymbol\theta^{(t)})\}$.
		Accordingly, we construct the following local quadratic surrogate functions 
		$\underline u_{r,\mathbf A}^{(t)}(\boldsymbol\theta)
		=
		u_{r,\mathbf A}(\boldsymbol\theta^{(t)})
		+
		\nabla u_{r,\mathbf A}^T(\boldsymbol\theta^{(t)})
		(\boldsymbol\theta-\boldsymbol\theta^{(t)})
		-
		\frac{\delta_{r,\mathbf A}^{(t)}}{2}
		\|\boldsymbol\theta-\boldsymbol\theta^{(t)}\|_2^2$
		and
		$\overline u_{r,\mathbf A}^{(t)}(\boldsymbol\theta)
		=
		u_{r,\mathbf A}(\boldsymbol\theta^{(t)})
		+
		\nabla u_{r,\mathbf A}^T(\boldsymbol\theta^{(t)})
		(\boldsymbol\theta-\boldsymbol\theta^{(t)})
		+
		\frac{\delta_{r,\mathbf A}^{(t)}}{2}
		\|\boldsymbol\theta-\boldsymbol\theta^{(t)}\|_2^2$,
		where $\delta_{r,\mathbf A}^{(t)}>0$ is the curvature parameter.
		The two surrogate functions are tight and gradient-consistent at
		$\boldsymbol\theta^{(t)}$. The curvature parameters
		$\delta_{r,\mathbf A}^{(t)}>0$ are introduced to control the local
		curvature of the quadratic approximations within the prescribed trust region.
		
		Accordingly, define $\mathbf W_k\triangleq\mathbf w_k\mathbf w_k^H$ and
		$\mathbf S\triangleq\sum_{k=1}^{K_{\mathrm I}}\mathbf W_k+\mathbf V$.
		The corresponding surrogate functions are denoted by
		$\underline U_{i,k}^{(t)}\triangleq\underline u_{i,\mathbf W_k}^{(t)}$,
		$\overline U_{i,k}^{(t)}\triangleq\overline u_{i,\mathbf W_k}^{(t)}$,
		$\underline U_{i,V}^{(t)}\triangleq\underline u_{i,\mathbf V}^{(t)}$,
		$\overline U_{i,V}^{(t)}\triangleq\overline u_{i,\mathbf V}^{(t)}$,
		$\underline E_{j,k}^{(t)}\triangleq\underline u_{j^{\mathrm E},\mathbf W_k}^{(t)}$,
		$\overline E_{j,k}^{(t)}\triangleq\overline u_{j^{\mathrm E},\mathbf W_k}^{(t)}$,
		$\underline E_{j,V}^{(t)}\triangleq\underline u_{j^{\mathrm E},\mathbf V}^{(t)}$,
		$\overline E_{j,V}^{(t)}\triangleq\overline u_{j^{\mathrm E},\mathbf V}^{(t)}$,
		and
		$\underline P_j^{(t)}\triangleq\underline u_{j^{\mathrm E},\mathbf S}^{(t)}$,
		where $\mathbf c_i=\mathbf h_i$ and $\mathbf c_{j^{\mathrm E}}=\mathbf g_j$.
		Then, problem~\eqref{p4} is approximated by the following convex surrogate problem
		\begin{subequations}\label{p5}
			\begin{align}
				\max_{\boldsymbol\theta}~
				&\sum\nolimits_{j=1}^{K_{\mathrm E}}
				\underline P_j^{(t)}(\boldsymbol\theta)
				\label{p5-a}\\
				{\rm s.t.}~
				&\frac{\underline U_{i,i}^{(t)}(\boldsymbol\theta)}
				{\gamma_{\min}^{\mathrm{IR}}}
				\geq
				\sum_{k\ne i}\overline U_{i,k}^{(t)}(\boldsymbol\theta)
				+\overline U_{i,V}^{(t)}(\boldsymbol\theta)
				+\sigma_i^2,\forall i,
				\label{p5-b}\\
				&\underline P_j^{(t)}(\boldsymbol\theta)
				\geq B_j,\forall j,
				\label{p5-c}\\
				&\frac{\overline E_{j,i}^{(t)}(\boldsymbol\theta)}
				{\gamma_{\max}^{\mathrm{ER}}}
				\leq
				\sum_{k\ne i}\underline E_{j,k}^{(t)}(\boldsymbol\theta)
				+\underline E_{j,V}^{(t)}(\boldsymbol\theta)
				+\sigma_j^2,\forall i,j,
				\label{p5-d}\\
				&0\leq\theta_{z,n}\leq\theta_{\max},\quad\forall n,
				\label{p5-e}\\
				&\|\boldsymbol\theta-\boldsymbol\theta^{(t)}\|_2
				\leq\Delta_{\rm trust}.
				\label{p5-f}
			\end{align}
		\end{subequations}
		
		The azimuth angles are optimized using unwrapped local coordinates within the trust region and are mapped to $[0,2\pi)$ after an accepted update.
		The objective function in \eqref{p5} is concave, and constraints
		\eqref{p5-b}--\eqref{p5-f} are convex. Hence, problem \eqref{p5} is a convex QCQP and can be solved using CVX.

	\subsection{The Proposed SCA-based AO Algorithm}
	
	\begin{algorithm}[!t]
		\caption{SCA-based AO Algorithm for Solving (\ref{p1})}
		\label{alg1}
		\begin{algorithmic}[1]
			\STATE Initialize the feasible variables $\boldsymbol\theta^{(0)}$, $\{\mathbf W_i^{(0)}\}$, and $\mathbf V^{(0)}$; set iteration index as $t = 0$.
			\STATE Reformulate problem (\ref{p1}) as problem (\ref{p2}).
			\REPEAT
			\STATE $t = t+1$.
			\STATE For given ${\bm \theta}^{(t-1)}$, update ${\bf W}^{(t)}$ and ${\bf V}^{(t)}$ by solving the SDR problem (\ref{p3}).
			\STATE Recover ${\bf w}^{(t)}$ based on \textit{Theorem 1}.
			\STATE For given ${\bf w}^{(t)}$, ${\bf V}^{(t)}$, and ${\bm \theta}^{(t-1)}$, update ${\bm \theta}^{(t)}$ by solving problem (\ref{p5}).
			\UNTIL Convergence.
			\STATE {Output:} ${\bm \theta} = {\bm \theta}^{(t)}$, ${\bf w} = {\bf w}^{(t)}$, and ${\bf V} = {\bf V}^{(t)}$.
		\end{algorithmic}
	\end{algorithm}
	The resulting procedure is summarized in {\bf Algorithm~\ref{alg1}}.
	Using the standard interior-point method, the computational complexity
	of one beamforming update is approximately $C_{\rm BF}
	=\mathcal O\left(
	\sqrt{\beta_{\rm BF}}
	(K_{\rm I}+1)^3N^6
	\ln\left(\frac{1}{\epsilon_{\rm BF}}\right)
	\right)$,
	where
	$\beta_{\rm BF}
	=K_{\rm I}+K_{\rm E}+K_{\rm I}K_{\rm E}+1+(K_{\rm I}+1)N$,
	and $\epsilon_{\rm BF}$ denotes the solution accuracy.
	For the SCA-based RA-rotation subproblem, the complexity of solving
	problem~(\ref{p5}) once is approximately $C_{\rm RA}
	=\mathcal O\left(
	8\sqrt{\beta_{\rm RA}}N^3
	\ln\left(\frac{1}{\epsilon_{\rm RA}}\right)
	\right)$,
	where
	$\beta_{\rm RA}
	=K_{\rm I}+K_{\rm E}+K_{\rm I}K_{\rm E}+2N+1$,
	and $\epsilon_{\rm RA}$ denotes the solution accuracy.
	Therefore, if $T_1$ AO iterations are required for convergence, the
	overall computational complexity is $\mathcal O\left(
	T_1(C_{\rm BF}+C_{\rm RA})
	\right)$.
	
	At each AO iteration, the beamforming subproblem is first optimized
	for the current RA orientations, followed by one SCA-based rotation
	update. The curvature parameters are selected to ensure that the
	constructed surrogate functions provide valid local lower and upper
	bounds within the prescribed trust region. Since the surrogate
	functions are tight at the current point, the updated solution remains
	feasible and does not decrease the original objective value, i.e.,
	$F^{(t)}\geq F^{(t-1)}$. Since the transmit power is bounded by $P_0$
	and the channel gains are finite, the objective function is upper
	bounded. Therefore, the sequence $\{F^{(t)}\}$ converges.

	\section{The Proposed DE-based AO Algorithm}
	
	In the previous section, we developed the SCA-based AO algorithm, which exploits the analytical structure of the considered problem through channel Jacobians and successive convex approximations. This approach provides a mathematically tractable framework, where each RA rotation update is obtained by solving a convex surrogate problem and the monotonicity and convergence of the objective sequence can be established. However, the computation of channel derivatives, construction of convex surrogate functions, and solution of the resulting convex subproblems may incur considerable computational complexity as the number of RAs increases.
	
	To provide an alternative solution strategy with simpler implementation, we further develop a DE-based AO algorithm. Different from the SCA-based method, DE does not require convexification of the RA rotation subproblem or gradient information. Instead, it directly searches over the continuous rotation-angle space through three basic operators, namely mutation, crossover, and selection \cite{storn1997differential}. Therefore, the nonlinear polarization-aware channel model can be handled through numerical channel and constraint evaluations without deriving explicit convex approximations. This feature also facilitates the use of numerical or measured antenna radiation patterns. As a population-based heuristic method, DE provides broader exploration of the highly non-convex rotation-angle space and has a simple algorithmic structure with only a few control parameters. Although its stochastic search does not guarantee convergence to a global or local optimum, the computational cost of its RA rotation search grows more mildly with the number of RAs, making it an attractive alternative for large-scale RA rotation optimization.
	
	\subsection{Differential Evolution Algorithm for Rotation Optimization}
	
	Following the AO framework in Section~\uppercase\expandafter{\romannumeral3}, the beamforming subproblem is solved in the same manner as in Section~\uppercase\expandafter{\romannumeral3}-A. For given $\mathbf W$ and $\mathbf V$, the RA rotation subproblem can be written as
	\begin{subequations}\label{p9}
		\begin{align}
			\mathop {\max }\limits_{\boldsymbol\theta}\quad 
			&\sum\nolimits_{j = 1}^{{K_E}} {{\rm{Tr}}\left[ {{\bf{G}}_j(\boldsymbol\theta)\Big( {\sum\nolimits_{i = 1}^{{K_I}} {{\bf{W}}_i}  + {\bf{V}}} \Big)} \right]}  \label{p9-a}\\
			{\rm s.t.}\quad 
			&\left(\rm \ref{p2-b}\right), \left(\rm \ref{p2-c}\right), \left(\rm \ref{p2-d}\right), \left(\rm \ref{p1-f}\right).
		\end{align}
	\end{subequations}
	
	Problem (\ref{p9}) is non-convex due to the nonlinear dependence of the channel matrices on $\boldsymbol\theta$. We therefore solve the rotation subproblem using DE, as described below.
	
	\subsubsection{Encoding of RA Rotation Variables}
	Since the direction of each RA is characterized by its zenith and azimuth angles, the DE individual is defined as 
	\begin{flalign}
		\bm{\theta}=\left[
		\theta_{z,1},\theta_{a,1},\ldots,\theta_{z,N},\theta_{a,N}
		\right]^{T}
		\in \mathbb{R}^{2N}.
	\end{flalign}
	For any $\bm \theta$, the pointing vector of RA $n$ is reconstructed by
	\begin{flalign}
		\vec{\bf{f}}_n(\theta_{z,n},\theta_{a,n})=
		\left[
		\sin\theta_{z,n}\cos\theta_{a,n},
		\sin\theta_{z,n}\sin\theta_{a,n},
		\cos\theta_{z,n}
		\right]^{T},
	\end{flalign}
	yielding the rotation set $\vec{\bf f} = \big\{\vec{\bf f}_n\big\}_{n=1}^N$.
	
	All rotation angles must satisfy the physical bounds
	\begin{flalign}
		0 \le \theta_{z,n} \le \theta _{\max}, ~~ 0 \le \theta_{a,n} < 2 \pi .\label{25}
	\end{flalign}
	
	These bounds are enforced during initialization and every DE update step.
	
	\subsubsection{Objective Function Evaluation}
	
	For any candidate rotation vector $\boldsymbol\theta$, the channel matrices
	$\mathbf H_i(\boldsymbol\theta)$ and $\mathbf G_j(\boldsymbol\theta)$ are
	obtained from \eqref{hni}--\eqref{gj}. For given $\mathbf W$ and $\mathbf V$,
	define $\mathbf S=\sum_i\mathbf W_i+\mathbf V$ and
	$U_j(\boldsymbol\theta)={\rm Tr}[\mathbf G_j(\boldsymbol\theta)\mathbf S]$.
	The objective value is then given by
	\begin{flalign}
		U(\boldsymbol\theta)
		=\sum\nolimits_{j=1}^{K_{\mathrm E}}U_j(\boldsymbol\theta).
		\label{eq37}
	\end{flalign}
	
	For each candidate orientation, the polarization-related quantities and
	channel coefficients are recalculated according to the model in
	Section II-A before evaluating \eqref{eq37}. A measured or standardized
	element pattern can also replace \eqref{Ge} without changing the DE procedure.
	
	\subsubsection{Handling the Non-convex Constraints}
	
	The rotation subproblem inherits the constraints in (\ref{p2-b})--(\ref{p2-d}) and the angle constraints in (\ref{p1-f}). To handle the non-convex functional constraints, DE employs a violation-based feasibility rule. To avoid combining constraint violations with different scales, the three functional constraints are normalized as
	\begin{flalign}
		&c_{1,i}=\frac{\gamma_{\min}^{\rm IR}-\gamma_i^{\rm IR}(\boldsymbol\theta)}
		{\gamma_{\min}^{\rm IR}}\leq0,\quad\forall i,\notag\\
		&c_{2,j}=\frac{B_j-U_j(\boldsymbol\theta)}{B_j}\leq0,\quad\forall j,\notag\\
		&c_{3,i,j}=\frac{\gamma_{i,j}^{\rm ER}(\boldsymbol\theta)-\gamma_{\max}^{\rm ER}}
		{\gamma_{\max}^{\rm ER}}\leq0,\quad\forall i,j.
	\end{flalign}
	
	The angle constraints in (\ref{p1-f}) are directly enforced during initialization and each DE update, and therefore need not be included in the constraint-violation function. The total constraint violation is defined as
	\begin{flalign}
		{\rm{CV}}\left(\bm{\theta}\right)
		&=\sum\nolimits_i\max\{0,c_{1,i}\}
		+\sum\nolimits_j\max\{0,c_{2,j}\}\notag\\
		&\quad+\sum\nolimits_{i,j}\max\{0,c_{3,i,j}\}. \label{eq39}
	\end{flalign}
	All terms in ${\rm CV}(\bm{\theta})$ are dimensionless.
	
	Selection follows Deb's feasibility rules: 1) if one solution is feasible and the other is infeasible, the feasible solution is selected; 2) if both are feasible, the solution with the larger $U(\boldsymbol\theta)$ is selected; and 3) if both are infeasible, the solution with the smaller ${\rm CV}(\boldsymbol\theta)$ is selected.
	
	\subsubsection{Differential Evolution Procedure}
	
	Let $N_{\rm pop}$ denote the population size and $G_{\max}$ the maximum number of generations. The DE procedure consists of mutation, crossover, and selection.
	
	{\textit{a) Mutation:}} For the $r$-th target vector at generation $g$, three distinct indices $r_1,r_2,r_3\ne r$ are selected. The classical \texttt{DE/rand/1} mutation generates
	\begin{flalign}
		{\bf v}_r^{(g)}
		=
		{\bm \theta}_{r_1}^{(g-1)}
		+
		F_g\left(
		{\bm \theta}_{r_2}^{(g-1)}
		-
		{\bm \theta}_{r_3}^{(g-1)}
		\right),
		\label{eq40}
	\end{flalign}
	where $F_g\in[F_{\min},F_{\max}]$ is the mutation factor. At each generation, $F_g$ is sampled from the prescribed interval. The zenith angles are projected onto $[0,\theta_{\max}]$, while the azimuth angles are wrapped modulo $2\pi$.
	
	{\textit{b) Crossover:}} A trial vector ${\bf u}_r^{(g)}$ is generated through binomial crossover as
	\begin{flalign}
		u_{r,d}^{(g)}
		=
		\left\{
		\begin{array}{ll}
			v_{r,d}^{(g)}, &
			{\rm if~rand}_d\leq {\rm CR}_g
			\ {\rm or}\ d=d_{\rm rand},\\
			\theta_{r,d}^{(g-1)}, &
			{\rm otherwise},
		\end{array}
		\right.
		\label{eq41}
	\end{flalign}
	where ${\rm CR}_g\in[{\rm CR}_{\min},{\rm CR}_{\max}]$ is the crossover rate, and $d_{\rm rand}$ ensures that at least one component is inherited from the mutant vector. The value of ${\rm CR}_g$ is sampled from its prescribed interval at each generation.
	
	{\textit{c) Selection:}} The target and trial vectors are compared according to Deb's feasibility rules. The preferred vector is retained in the next generation. The best feasible individual is preserved during the evolution process.
	
	\subsection{The Proposed DE-based AO Algorithm}
	
	Based on the DE procedure described above, we further integrate the rotation search with the beamforming optimization in an AO framework. The RA angles are initialized as $\boldsymbol\theta^{(0)}=\mathbf 0$. In each AO iteration, the beamforming variables are first optimized for the current RA orientations according to Section~\uppercase\expandafter{\romannumeral3}-A. Then, with the updated beamforming variables fixed, the DE algorithm is employed to optimize the RA rotation angles, where $\boldsymbol\theta^{(t-1)}$ is included in the initial population and the best feasible individual is adopted as $\boldsymbol\theta^{(t)}$. The overall procedure is summarized in {\bf Algorithm~\ref{alg2}}.
	
	\begin{algorithm}[!t]
		\caption{DE-based AO Algorithm for Solving (\ref{p1})}
		\label{alg2}
		\begin{algorithmic}[1]
			\STATE Initialize the feasible variables $\boldsymbol\theta^{(0)}$,
			$\{\mathbf W_i^{(0)}\}$, and $\mathbf V^{(0)}$; set $t=0$.
			\STATE Reformulate problem (\ref{p1}) as problem (\ref{p2}).
			\REPEAT
			\STATE $t=t+1$.
			\STATE For given $\boldsymbol\theta^{(t-1)}$, update
			$\{\mathbf W_i^{(t)}\}$ and $\mathbf V^{(t)}$ by solving the SDR of problem (\ref{p3}).
			\STATE Recover $\{\mathbf w_i^{(t)}\}$ based on \textit{Theorem~1}.
			\STATE Initialize the DE population for problem (\ref{p9}) with
			$\boldsymbol\theta^{(t-1)}$ included; set $g=0$.
			\REPEAT
			\STATE $g=g+1$.
			\STATE Generate the mutant vectors according to (\ref{eq40}).
			\STATE Generate the trial vectors according to (\ref{eq41}).
			\STATE Enforce the angle constraints and evaluate $U(\boldsymbol\theta)$
			and $CV(\boldsymbol\theta)$ according to (\ref{eq39}).
			\STATE Perform selection according to Deb's feasibility rules.
			\UNTIL $g=G_{\max}$.
			\STATE Update $\boldsymbol\theta^{(t)}$ with the best feasible individual.
			\UNTIL $t=T_{\max}$.
			\STATE {Output:} $\boldsymbol\theta=\boldsymbol\theta^{(t)}$,
			$\mathbf w=\mathbf w^{(t)}$, and $\mathbf V=\mathbf V^{(t)}$.
		\end{algorithmic}
	\end{algorithm}
	
	Let
	$C_{\rm eval}
	=\mathcal O\left(
	N + K_E + K_I K_E
	\right)$
	denote the computational cost of constructing the
	polarization-aware multipath channels and evaluating the objective
	function and constraints for one DE individual. Moreover, let
	$C_{\rm BF}$ denote the computational cost of one beamforming update,
	as defined in Section~\uppercase\expandafter{\romannumeral3}-C.
	In each AO iteration, the DE algorithm performs $G_{\max}$ generations
	with a population size of $N_{\rm pop}$. Therefore, the overall
	computational complexity of the DE-based AO algorithm is $\mathcal O\left(
	T_{\max}\left[
	C_{\rm BF}
	+G_{\max}N_{\rm pop}C_{\rm eval}
	\right]
	\right)$.
	
	At the $t$-th AO iteration, $\boldsymbol\theta^{(t-1)}$ is included
	in the initial DE population and the best feasible individual is
	preserved throughout the evolution process. Hence, for the optimized
	$\{\mathbf W_i^{(t)}\}$ and $\mathbf V^{(t)}$, the DE update cannot
	decrease the objective value with respect to
	$\boldsymbol\theta^{(t-1)}$. Since the beamforming subproblem is also
	optimized for each given RA orientation, the objective sequence
	$\{F^{(t)}\}$ is monotonically non-decreasing. Moreover, the objective
	is upper bounded due to the finite channel gains and the transmit-power
	constraint $P_0$. Therefore, the objective sequence converges.
	
		\section{Robust Beamforming Under Imperfect CSI}
		The preceding algorithms assume nominal effective CSI. To account for
		practical channel estimation errors, we further consider a worst-case
		robust beamforming design. Specifically, the RA orientations are first
		optimized using {\bf Algorithm~\ref{alg1}} or {\bf Algorithm~\ref{alg2}}
		based on the estimated channels and then fixed. The information
		beamforming and AN covariance matrices are subsequently re-optimized
		under the norm-bounded uncertainty model
		$\mathbf h_i=\widehat{\mathbf h}_i+\Delta\mathbf h_i$ and
		$\mathbf g_j=\widehat{\mathbf g}_j+\Delta\mathbf g_j$, where
		$\|\Delta\mathbf h_i\|_2\leq\varepsilon_i^{\mathrm I}$ and
		$\|\Delta\mathbf g_j\|_2\leq\varepsilon_j^{\mathrm E}$.
		Define
		$\mathbf S\triangleq\sum_i\mathbf W_i+\mathbf V$,
		$\mathbf Q_i\triangleq\mathbf W_i-\gamma_{\min}^{\mathrm{IR}}
		(\sum_{k\ne i}\mathbf W_k+\mathbf V)$, and
		$\mathbf R_{i,j}\triangleq\gamma_{\max}^{\mathrm{ER}}
		(\sum_{k\ne i}\mathbf W_k+\mathbf V)-\mathbf W_i$.
		Moreover, $\tau_j$ denotes the guaranteed worst-case received power
		at ER $j$. The robust beamforming problem is formulated as
		\begin{subequations}\label{p10}
			\begin{align}
				&\max_{\left\{{\mathbf W}_i\succeq\mathbf0\right\},\mathbf V\succeq\mathbf0,\boldsymbol\tau}\quad
				\sum\nolimits_j\tau_j \label{p10-a}\\
				\!\!{\rm s.t.}
				&\min_{\|\Delta\mathbf h_i\|\leq\varepsilon_i^{\mathrm I}}
				(\widehat{\mathbf h}_i\!+\!\Delta\mathbf h_i)^H\mathbf Q_i(\widehat{\mathbf h}_i\!+\!\Delta\mathbf h_i)
				\geq\gamma_{\min}^{\mathrm{IR}}\sigma_i^2,\forall i, \label{p10-b}\\
				&\min_{\|\Delta\mathbf g_j\|\leq\varepsilon_j^{\mathrm E}}
				(\widehat{\mathbf g}_j\!+\!\Delta\mathbf g_j)^H\mathbf S(\widehat{\mathbf g}_j\!+\!\Delta\mathbf g_j)
				\geq\tau_j,\forall j,\label{p10-c}\\
				&\tau_j\geq B_j,\ \forall j,\label{p10-d}\\
				&\min_{\|\Delta\mathbf g_j\|\leq\varepsilon_j^{\mathrm E}}
				(\widehat{\mathbf g}_j\!+\!\Delta\mathbf g_j)^H\mathbf R_{i,j}(\widehat{\mathbf g}_j\!+\!\Delta\mathbf g_j)
				\!+\!\gamma_{\max}^{\mathrm{ER}}\sigma_j^2\!\geq\! 0,\forall i,j,\label{p10-e}\\
				&\sum\nolimits_i\mathrm{Tr}(\mathbf W_i)+\mathrm{Tr}(\mathbf V)\leq P_0,\label{p10-f}\\
				&{\rm rank}(\mathbf W_i)=1,\quad\forall i.\label{p10-g}
			\end{align}
		\end{subequations}
		
		Since the uncertainty sets have nonempty interiors, the semi-infinite constraints in problem~\eqref{p10} can be equivalently transformed into finite LMIs via the S-Procedure \cite{4838902,yang2026rotatable}. In particular, constraint~\eqref{p10-b} is equivalent to the existence of $\lambda_i\geq0$ such that
		\begin{flalign}
			\begin{bmatrix}
				\mathbf Q_i+\lambda_i\mathbf I_N
				&
				\mathbf Q_i\widehat{\mathbf h}_i\\
				\widehat{\mathbf h}_i^H\mathbf Q_i
				&
				\widehat{\mathbf h}_i^H\mathbf Q_i\widehat{\mathbf h}_i
				-\gamma_{\min}^{\mathrm{IR}}\sigma_i^2
				-\lambda_i(\varepsilon_i^{\mathrm I})^2
			\end{bmatrix}
			\succeq\mathbf0.
			\label{robust_IR_LMI}
		\end{flalign}
		Similarly, constraint~\eqref{p10-c} is equivalent to the existence
		of $\mu_j\geq0$ such that
		\begin{flalign}
			\begin{bmatrix}
				\mathbf S+\mu_j\mathbf I_N
				&
				\mathbf S\widehat{\mathbf g}_j\\
				\widehat{\mathbf g}_j^H\mathbf S
				&
				\widehat{\mathbf g}_j^H\mathbf S\widehat{\mathbf g}_j
				-\tau_j-\mu_j(\varepsilon_j^{\mathrm E})^2
			\end{bmatrix}
			\succeq\mathbf0.
			\label{robust_EH_LMI}
		\end{flalign}
		The worst-case eavesdropping constraint in~\eqref{p10-e} is
		equivalent to the existence of $\nu_{i,j}\geq0$ such that
		\begin{flalign}
			\begin{bmatrix}
				\mathbf R_{i,j}\!+\!\nu_{i,j}\mathbf I_N & \mathbf R_{i,j}\widehat{\mathbf g}_j\\
				\widehat{\mathbf g}_j^H\mathbf R_{i,j} &
				\widehat{\mathbf g}_j^H\mathbf R_{i,j}\widehat{\mathbf g}_j
				\!+\!\gamma_{\max}^{\mathrm{ER}}\sigma_j^2\!-\!\nu_{i,j}(\varepsilon_j^{\mathrm E})^2
			\end{bmatrix}\succeq\mathbf0. \label{robust_ER_LMI}
		\end{flalign}
		
		After replacing the semi-infinite constraints with
		\eqref{robust_IR_LMI}--\eqref{robust_ER_LMI} and relaxing the
		rank-one constraints in~\eqref{p10-g}, the resulting problem becomes
		an SDP and can be solved using CVX. If the obtained information
		covariance matrices are rank one, the beamforming vectors can be
		directly recovered. Otherwise, Gaussian randomization is applied to
		generate candidate beamforming directions. For each candidate, the
		transmit powers, AN covariance matrix, and S-Procedure multipliers are re-optimized subject to the robust LMIs. The feasible candidate with the largest worst-case sum power is selected.
	
	\section{Numerical Results}
	\begin{figure}[!t]
		\centering
		\includegraphics[width=2.5in]{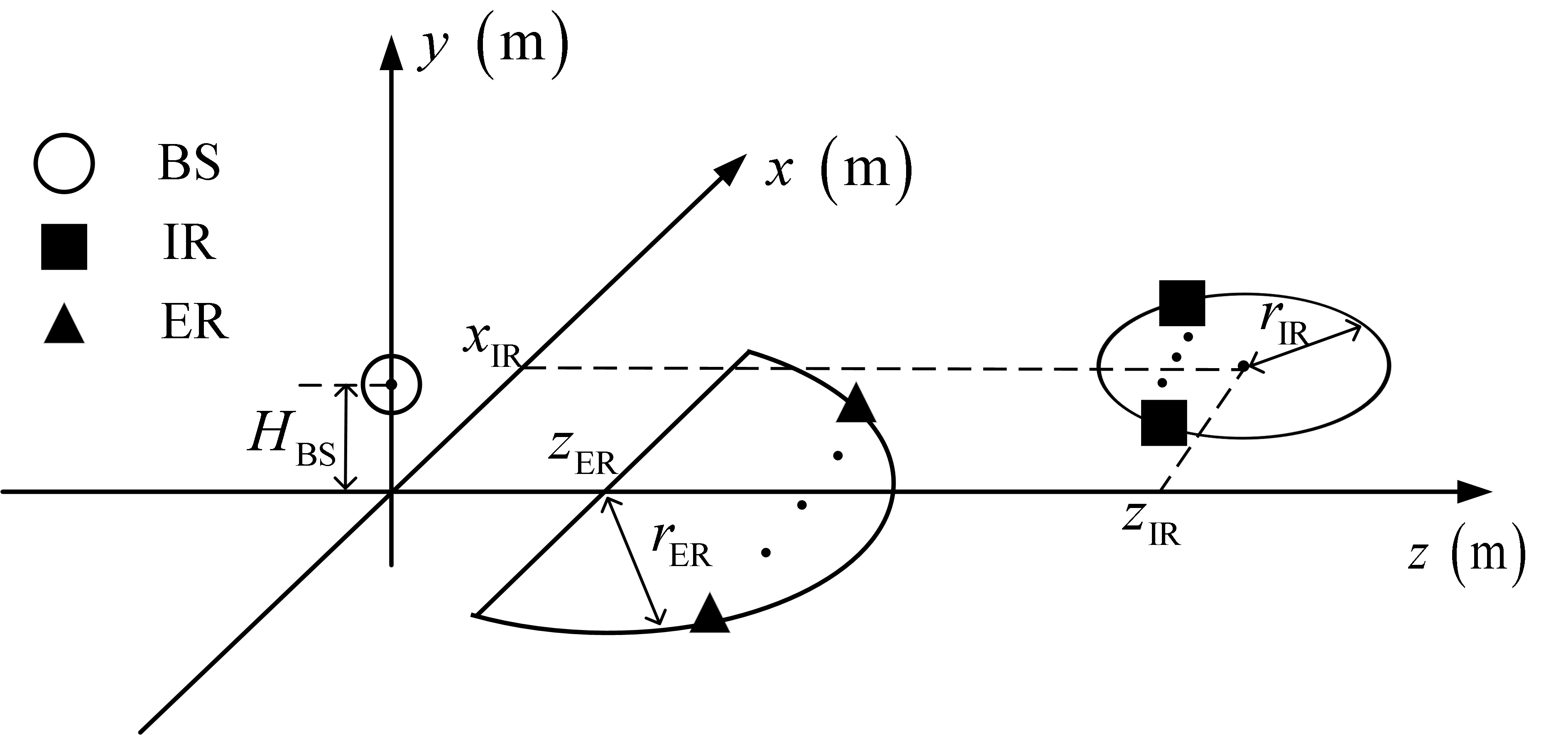}
		\caption{3D model of RA-aided secure SWIPT system.}
		\label{3D_model}
	\end{figure}
	This section presents the simulation results of the proposed system and algorithms. The corresponding three-dimensional simulation geometry is illustrated in Fig.~\ref{3D_model}. Following \cite{11427014}, we set $f_c=2.4~{\rm GHz}$, $\lambda=0.125~{\rm m}$, and $\Delta=\lambda/2$. Fig. \ref{3D_model} depicts the 3D coordinate system of the RA-aided secure SWIPT system. The BS is located at $(0,H_{\rm BS},0)$. The IRs are uniformly distributed on a circle centered at $\left(x_{\rm IR},0,z_{\rm IR}\right)$ with radius $r_{\rm IR}$, while the ERs are uniformly distributed on a semicircle centered at $\left(x_{\rm ER},0,z_{\rm ER}\right)$ with radius $r_{\rm ER}$, where $H_{\rm BS} = 3~{\rm m}$, $x_{\rm IR} = 4~{\rm m}$, $z_{\rm IR} = 20~{\rm m}$, $x_{\rm ER} = 0~{\rm m}$, $z_{\rm ER} = 1~{\rm m}$, $r_{\rm IR} = 3~{\rm m}$, and $r_{\rm ER} = 6~{\rm m}$. Unless otherwise specified, $N_x=N_y=3$, $N=9$, $K_{\rm I}=K_{\rm E}=2$, $\sigma_i^2=\sigma_j^2=-80~{\rm dBm}$, $\gamma_{\min}^{\rm IR}=10~{\rm dB}$, $\gamma_{\max}^{\rm ER}=0~{\rm dB}$, $\varepsilon_i^{\mathrm I} = \varepsilon_j^{\mathrm E} = 0.01$, $A_0=-30~{\rm dB}$, $P_0=40~{\rm dBm}$, $P_{\min}=-26~{\rm dBm}$, $\Delta_{\rm trust}=\pi/8$, $A = \lambda ^2 / 4 \pi$, $r_{\rm RA} = \lambda / \sqrt{8 \pi}$, $d_{\rm safe} = 0.1 \lambda$, $a_j=150$, $b_j=0.014$, $Z_j=0.024$, $p=4$, and $\theta_{\max}=60^\circ$, while all path-loss exponents are set to $2$. Polarization mismatch is considered, with the polarization directions of all IR and ER antennas fixed along the $y$-axis. Moreover, $Q=4$ scatterers are generated in the propagation environment to form the single-bounce NLoS paths. The reflection power coefficients are within $[-15,-5]~{\rm dB}$, while the scatterer horizontal distances and heights are generated within $[2,8]~{\rm m}$ and $[0.5,3]~{\rm m}$, respectively. For the DE-based AO algorithm, $N_{\rm pop}=100$, $G_{\max}=100$, $F_g\in[0.4,0.9]$, ${\rm CR}_g\in[0.6,0.95]$, and $T_{\max}=20$. All results are averaged over $50$ independent channel realizations with independently generated scatterers and reflection coefficients.
	
	\subsection{Comparison of the Two Proposed Algorithms}
	In this subsection, we examine the convergence characteristics and achievable performance of the proposed algorithms. To facilitate the subsequent presentation and comparison, two algorithmic schemes are introduced as follows: 1) {\bf SCA-based AO algorithm}: The proposed SCA-based AO method is used to optimize problem (\ref{p1}). 2) {\bf DE-based AO algorithm}: The proposed DE-based AO method is adopted to optimize problem (\ref{p1}).
	
	\begin{figure}[!t]
		\centering
		\includegraphics[width=2.3in]{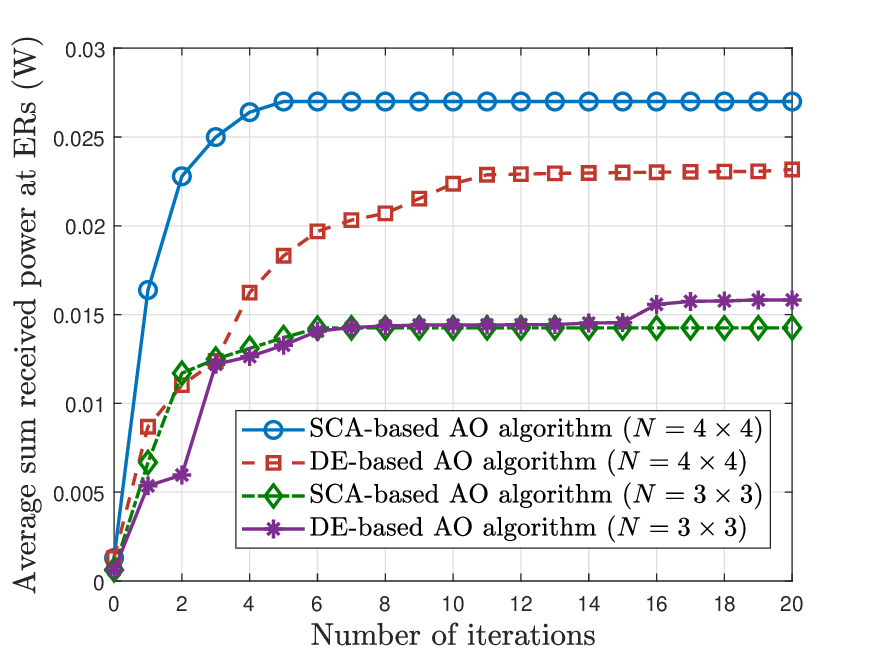}
		\caption{Convergence behavior of the proposed algorithms.}
		\label{iteration}
	\end{figure}
	Fig. \ref{iteration} illustrates the convergence behavior of the proposed SCA-based and DE-based AO algorithms. It can be observed that the SCA-based AO algorithm converges rapidly for both $N=3\times3$ and $N=4\times4$, with the objective value stabilizing after only a few AO iterations. In comparison, the DE-based AO algorithm requires more iterations to achieve stable performance. For $N=3\times3$, the DE-based AO algorithm eventually achieves a higher average sum received power than the SCA-based AO algorithm, benefiting from its population-based search capability. However, when $N$ increases to $4\times4$, the DE-based AO algorithm exhibits inferior performance under the same algorithm settings. This is because the increased number of RAs enlarges the dimension of the rotation-angle search space, which makes it more difficult for DE to sufficiently explore the feasible region with a fixed population size and number of generations. In addition, increasing $N$ significantly improves the sum received power due to the enhanced spatial beamforming gain.
	
	\begin{figure}[!t]
		\centering
		\includegraphics[width=2.3in]{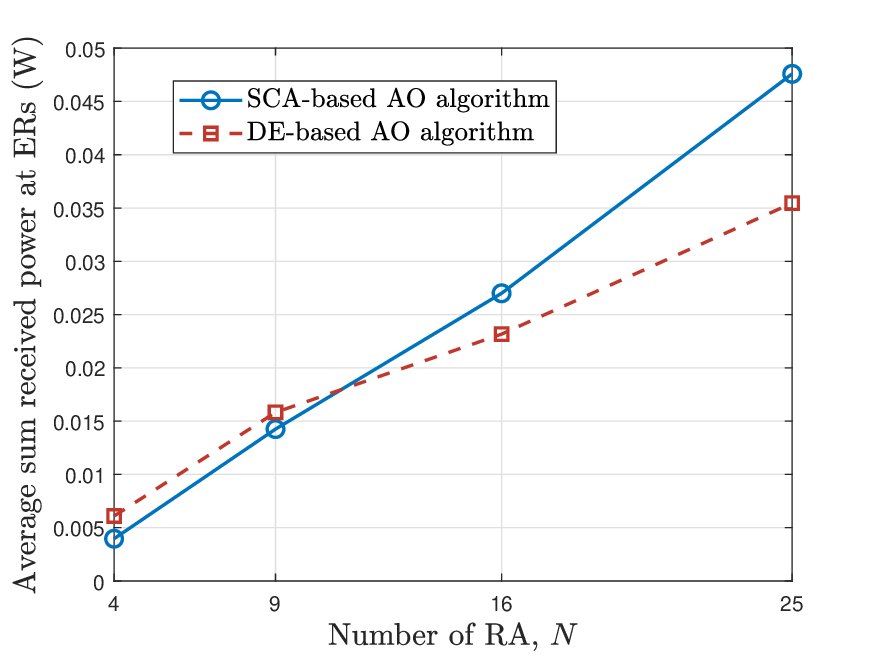}
		\caption{Performance of the SCA-based AO algorithm versus DE-based AO algorithm.}
		\label{vs}
	\end{figure}
	Fig.~\ref{vs} compares the performance of the proposed SCA-based and DE-based AO algorithms versus the number of RAs. When $N$ is small, the DE-based AO algorithm achieves a slightly higher average sum received power than the SCA-based AO algorithm. However, as $N$ increases, the SCA-based AO algorithm gradually outperforms the DE-based AO algorithm, and the performance gap becomes more pronounced. This is mainly because increasing $N$ enlarges the dimension of the RA rotation-angle search space, which makes it more difficult for the population-based DE algorithm to sufficiently explore the solution space under a fixed population size and number of generations. In contrast, the SCA-based AO algorithm updates the RA rotations through successive convex approximations and therefore maintains better performance as the problem dimension increases.
	
	\begin{figure}[!t]
		\centering
		\includegraphics[width=2.3in]{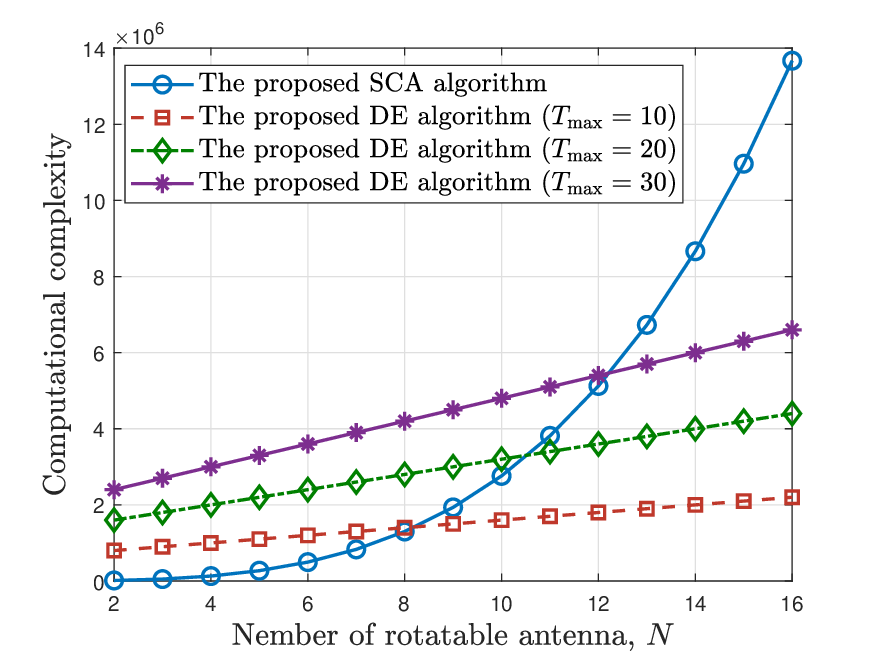}
		\caption{Comparison of the computational complexity for solving the RA rotation subproblem.}
		\label{comp}
	\end{figure}
	Fig.~\ref{comp} compares the dominant algorithm-specific computational costs of the proposed SCA-based and DE-based AO algorithms versus the number of RAs. Both algorithms share the same beamforming cost $C_{\rm BF}$, while their main difference lies in the RA rotation optimization. Specifically, the SCA-based method solves a convex surrogate problem with complexity $C_{\rm RA}$, whereas the DE-based method performs population-based search with a cost proportional to $G_{\max}N_{\rm pop}C_{\rm eval}$ per AO iteration. For a small number of RAs, the SCA-based rotation update has lower computational cost. As $N$ increases, however, the SCA-based rotation cost grows faster, while the DE-based search increases more mildly under fixed $G_{\max}$ and $N_{\rm pop}$. Increasing $T_{\max}$ also approximately proportionally increases the total cost of the DE-based AO algorithm. Together with Fig.~\ref{vs}, these results show that the two algorithms exhibit different performance and computational characteristics.

	\subsection{The Advantages of the RA systems}
	For a thorough performance comparison, the proposed RA-enabled secure SWIPT system is evaluated against the following benchmark schemes in this subsection. 
	1) {\bf The proposed scheme}: This refers to the considered RA-assisted secure SWIPT design, in which each RA can adjust its orientation independently. The proposed SCA-based AO algorithm is adopted to jointly optimize the BS beamforming and the RA rotation angles, thereby achieving stable convergence and high solution quality. 
	2) {\bf Array-wise orientation scheme}: In this case, all RAs are forced to share an identical rotation angle. This scheme can be viewed as a low-complexity RA-based design and is likewise solved by the proposed SCA-based AO algorithm. 
	3) {\bf Random orientation scheme}: In this benchmark, the rotation angles of all RAs are generated independently in a random manner. 
	4) {\bf Isotropic antenna scheme}: Under this setup, $N$ isotropic antennas are deployed at the BS for secure SWIPT transmission. 
	5) {\bf Fixed antenna scheme}: In this scheme, $N$ directional antennas are equipped at the BS, while their pointing directions remain unchanged during the whole transmission process.
	
	\begin{figure}[!t]
		\centering
		\includegraphics[width=2.3in]{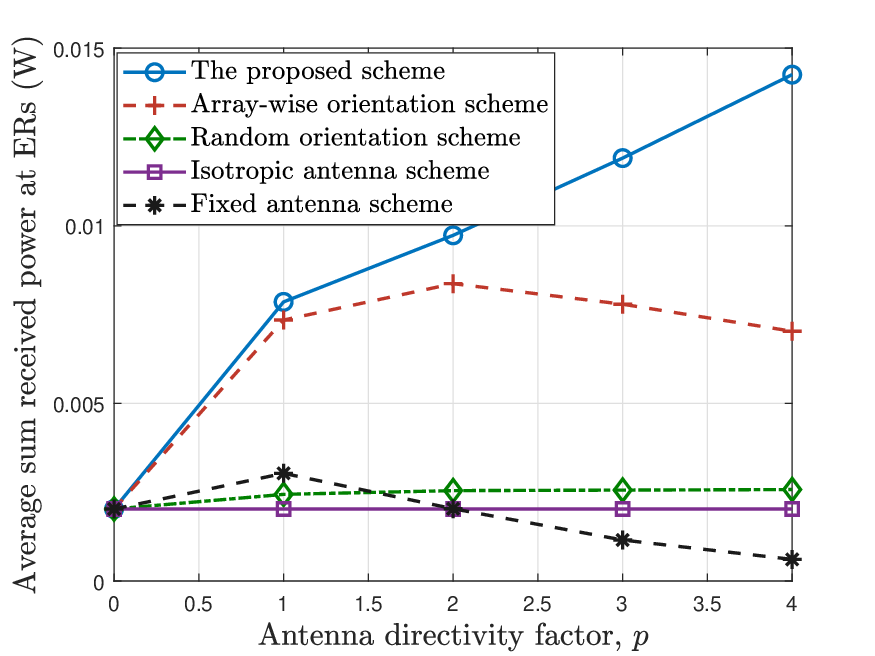}
		\caption{Average sum received power at ERs versus the antenna directivity factor.}
		\label{p}
	\end{figure}
	Fig. \ref{p} shows the sum received power at the ERs under different values of the antenna directivity factor. The performance of the proposed RA scheme improves with increasing $p$, since a larger $p$ yields a more concentrated radiation pattern along the pointing direction of the RAs, allowing stronger signal power to be delivered to the desired directions through RA rotation. In contrast, the performance of the array-wise orientation and the fixed antenna schemes first improves and then degrades as $p$ increases. This is because, in both schemes, the signal energy is primarily focused along the fixed boresight direction of the UPA. While increasing $p$ enhances the radiation strength in the intended direction, an excessively large $p$ significantly weakens signal coverage in other directions, making it difficult to serve non-aligned users and thereby degrading the overall system performance. The isotropic antenna scheme slightly outperforms the fixed antenna scheme, since isotropic antennas are able to provide relatively uniform radiation power to devices located in different directions. In contrast, the fixed antenna scheme directs most of its radiation power toward a predetermined direction, which limits its ability to effectively serve devices distributed over other directions.
	
	\begin{figure}[!t]
		\centering
		\includegraphics[width=2.3in]{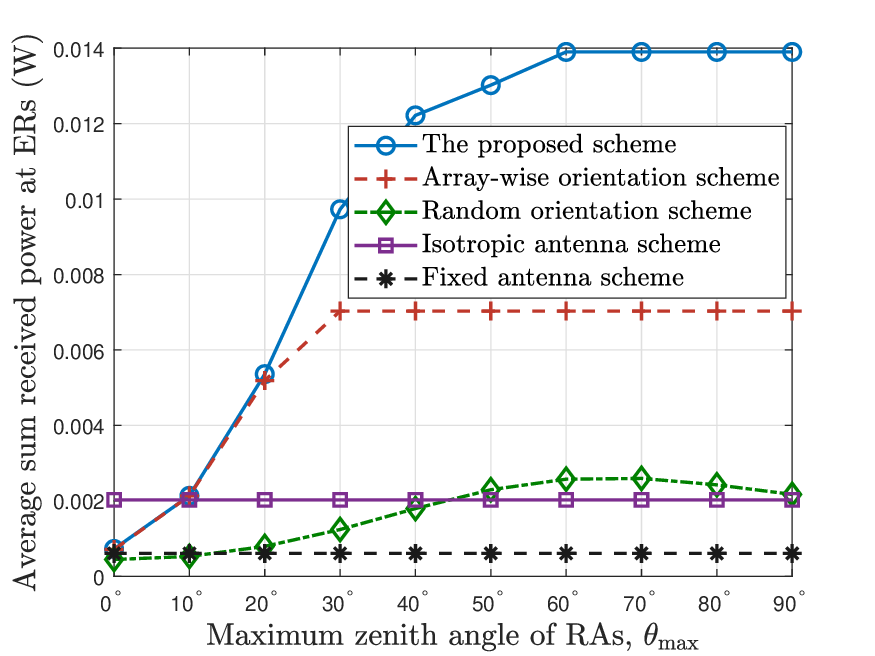}
		\caption{ Average sum received power at ERs versus the maximum zenith angle of RAs.}
		\label{theta}
	\end{figure}
	Fig. \ref{theta} shows the sum received power at the ERs for different maximum zenith angles of the RAs. We can observe that the performance of both the proposed RA scheme and the array-wise orientation scheme improves as the maximum zenith angle increases and gradually saturates when the angle reaches approximately $60^{\circ}$. This is because a larger zenith angle enables a wider boresight coverage range of the RAs, allowing the directional beams to be steered toward a broader set of IRs and ERs. Benefiting from its enhanced spatial DoFs, the proposed RA scheme consistently achieves better performance than the benchmark schemes. By contrast, the random orientation scheme exhibits a non-monotonic trend as the maximum zenith angle increases. A larger $\theta_{\max}$ provides a wider range of possible antenna orientations, which may initially increase the probability of favorable pointing directions. However, an excessively large rotation range increases the randomness of the antenna orientations, thereby degrading performance.
	
	\begin{figure}[!t]
		\centering
		\includegraphics[width=2.3in]{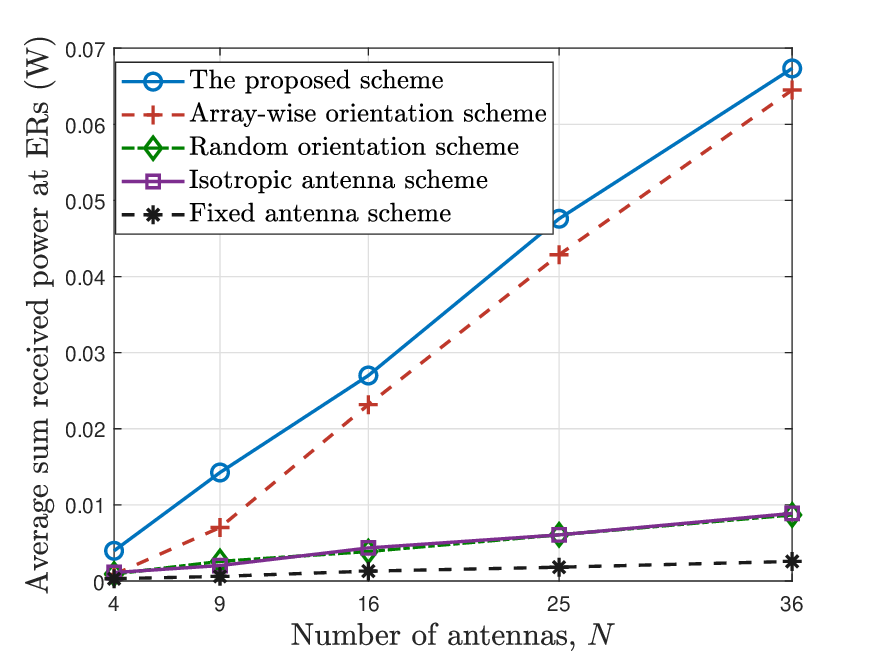}
		\caption{Average sum received power at ERs versus the number of RAs.}
		\label{N}
	\end{figure}
	Fig.~\ref{N} depicts the average sum received power at the ERs versus the number of antennas $N$. It can be observed that the received power increases with $N$ for all schemes, since deploying more antennas provides higher spatial beamforming gains and more degrees of freedom for energy transmission. Among all the considered schemes, the proposed RA design consistently achieves the best performance. In particular, it outperforms the array-wise orientation scheme because the independent rotation of each RA provides greater flexibility in adapting the antenna orientations to different propagation directions. In contrast, the random orientation and isotropic antenna schemes achieve substantially lower received power, while the fixed antenna scheme exhibits the worst performance due to its lack of orientation adaptation. These results demonstrate that jointly exploiting directional antenna gains and independent RA rotation becomes increasingly beneficial as the array size grows.
	
	\begin{figure}[!t]
		\centering
		\includegraphics[width=2.3in]{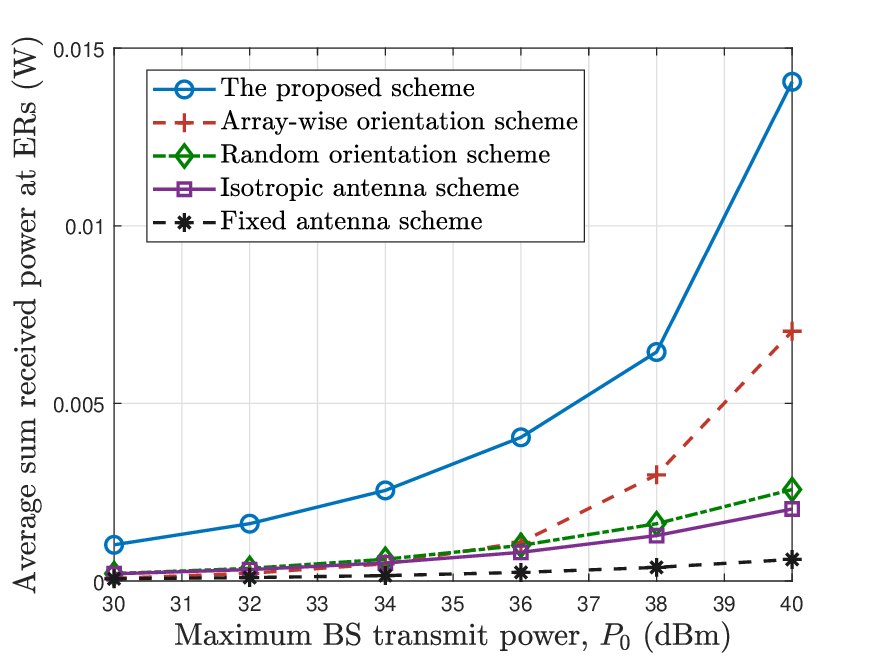}
		\caption{Average sum received power at ERs versus the transmit power at BS.}
		\label{P0}
	\end{figure}
	Fig.~\ref{P0} depicts the average sum received power at the ERs versus the maximum BS transmit power $P_0$. It can be observed that the received power increases with $P_0$ for all schemes, since a larger transmit power budget provides more available power for satisfying the information and energy transmission requirements. Among all the considered schemes, the proposed RA design consistently achieves the highest received power and its performance advantage becomes more pronounced as $P_0$ increases. This is because the independently optimized RA orientations can better exploit the directional antenna gains and spatial degrees of freedom to steer the additional transmit power toward the ERs. In comparison, the array-wise orientation scheme provides less orientation flexibility, while the random, isotropic, and fixed antenna schemes exhibit substantially lower performance. These results further demonstrate the benefit of jointly optimizing the BS beamforming and individual RA orientations, particularly in the high-transmit-power regime.
	
	\begin{figure}[!t]
		\centering
		\includegraphics[width=2.3in]{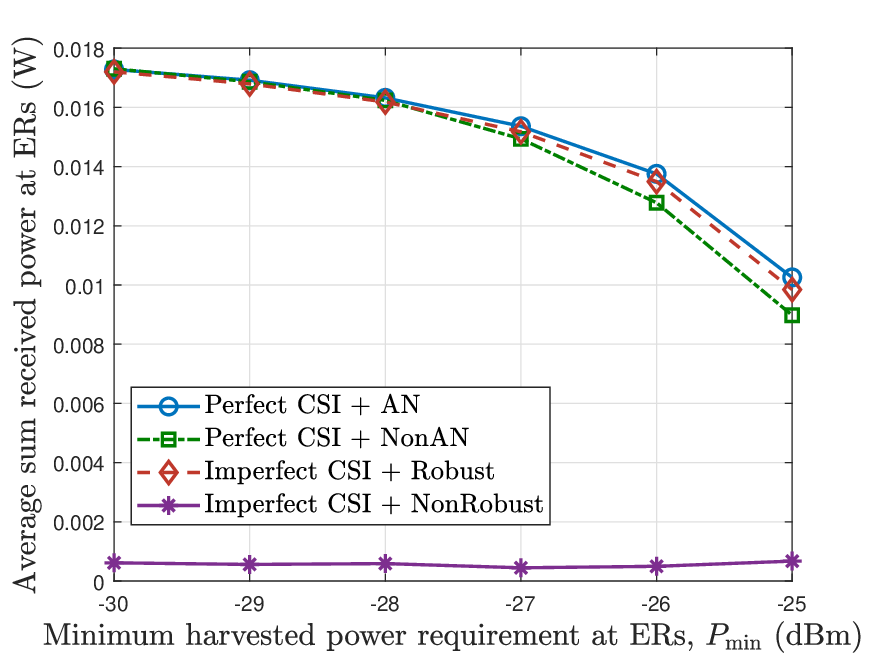}
		\caption{Average sum received power at the ERs versus the minimum harvested power requirement.}
		\label{Pmin}
	\end{figure}
	Fig.~\ref{Pmin} compares the average sum received power at the ERs under different CSI and beamforming designs versus the minimum harvested power requirement $P_{\min}$. It can be observed that the received power gradually decreases as $P_{\min}$ increases, since a more stringent EH requirement reduces the feasible design space and forces more transmit resources to satisfy the individual energy constraints. Under perfect CSI, the scheme with AN consistently outperforms the NoAN scheme because AN not only suppresses information leakage but also contributes to the power received at the ERs. Under imperfect CSI, the proposed robust beamforming design achieves performance close to that obtained under perfect CSI, demonstrating its effectiveness against channel estimation errors. In contrast, the non-robust design suffers a substantial performance loss because the beamforming solution optimized for the estimated channels cannot adequately account for CSI uncertainty. These results demonstrate the importance of both AN transmission and robust beamforming for secure SWIPT systems with imperfect CSI.

	\section{Conclusion}
	In this paper, we investigated an RA-enabled secure SWIPT system, where the spatial DoFs provided by RA rotation were exploited to enhance wireless power transfer while ensuring communication security. We formulated a joint BS beamforming and RA rotation design to maximize the sum received power at the ERs. To solve the resulting non-convex problem, an AO framework was developed, where the beamforming subproblem was handled via SDR with deterministic rank-one recovery, while the RA rotation subproblem was addressed by SCA- and DE-based methods. A worst-case robust beamforming extension was further developed for norm-bounded CSI errors. Simulation results demonstrated the performance advantages of the proposed RA-enabled design over fixed- and isotropic-antenna benchmarks, while the performance-oriented SCA-based and complexity-oriented DE-based algorithms exhibited distinct advantages in solution quality and computational efficiency, respectively. Moreover, the proposed robust beamforming design exhibited strong resilience to CSI uncertainties.

	\appendices
	
	\section{Proof of Theorem 1}
	
	Let $\{\widetilde{\mathbf W}_i\}_{i=1}^{K_I}$ and
	$\widetilde{\mathbf V}$ denote an arbitrary optimal solution to the
	{SDR of problem~\eqref{p3}}. Since $\gamma_{\min}^{\rm IR}>0$ and
	$\sigma_i^2>0$, constraint~\eqref{p3-b} implies that
	$\mathbf h_i^H(\bm \theta)\widetilde{\mathbf W}_i\mathbf h_i(\bm \theta)>0$, $\forall i$.
	Accordingly, we construct
	\begin{flalign}
		\mathbf w_i^\star
		\triangleq
		\frac{\widetilde{\mathbf W}_i\mathbf h_i(\bm \theta)}
		{\sqrt{\mathbf h_i^H(\bm \theta)\widetilde{\mathbf W}_i\mathbf h_i(\bm \theta)}},
		\mathbf W_i^\star
		\triangleq
		\mathbf w_i^\star(\mathbf w_i^\star)^H,
		\quad \forall i.
		\label{rank_one_construction}
	\end{flalign}
	It follows that $\mathbf w_i^\star\neq\mathbf 0$ and thus
	$\operatorname{rank}(\mathbf W_i^\star)=1$, $\forall i$.
	
	Define
	$\mathbf D_i\triangleq
	\widetilde{\mathbf W}_i-\mathbf W_i^\star$.
	For any $\mathbf x\in\mathbb C^N$, we have
	\begin{flalign}
		\mathbf x^H\mathbf D_i\mathbf x
		=
		\mathbf x^H\widetilde{\mathbf W}_i\mathbf x
		-
		\frac{
			\left|
			\mathbf x^H\widetilde{\mathbf W}_i\mathbf h_i(\bm \theta)
			\right|^2
		}{
			\mathbf h_i^H(\bm \theta)\widetilde{\mathbf W}_i\mathbf h_i(\bm \theta)
		}
		\geq 0,
		\label{residual_psd}
	\end{flalign}
	where the inequality follows from the Cauchy--Schwarz inequality.
	Hence, $\mathbf D_i\succeq\mathbf 0$. Moreover, according to
	\eqref{rank_one_construction},
	$\mathbf h_i^H(\bm \theta)\mathbf W_i^\star\mathbf h_i(\bm \theta)
	=
	\mathbf h_i^H(\bm \theta)\widetilde{\mathbf W}_i\mathbf h_i(\bm \theta)$, which implies
	$\mathbf h_i^H(\bm \theta)\mathbf D_i\mathbf h_i(\bm \theta)=0$.
	
	We further construct the AN covariance matrix as
	\begin{flalign}
		\mathbf V^\star
		\triangleq
		\widetilde{\mathbf V}
		+
		\sum_{i=1}^{K_I}\mathbf D_i,
		~
		\sum_{i=1}^{K_I}\mathbf W_i^\star+\mathbf V^\star
		=
		\sum_{i=1}^{K_I}\widetilde{\mathbf W}_i
		+\widetilde{\mathbf V}.
		\label{total_covariance_preserved}
	\end{flalign}
	Since $\widetilde{\mathbf V}\succeq\mathbf 0$ and
	$\mathbf D_i\succeq\mathbf 0$, we have
	$\mathbf V^\star\succeq\mathbf 0$. Furthermore,
	\eqref{total_covariance_preserved} shows that the total transmit
	covariance matrix remains unchanged. Therefore, the objective value,
	the total transmit-power constraint, and the received power
	constraints at all ERs remain unchanged.
	
	Next, we verify the IR SINR constraints. The desired signal power of
	IR $i$ remains unchanged, while its interference-plus-AN power
	satisfies
	\begin{flalign}
		&\sum\nolimits_{k\neq i}
		\mathbf h_i^H(\bm \theta)\mathbf W_k^\star\mathbf h_i(\bm \theta)
		+
		\mathbf h_i^H(\bm \theta)\mathbf V^\star\mathbf h_i(\bm \theta)
		\notag\\
		&~~{}=
		\sum\nolimits_{k\neq i}
		\mathbf h_i^H(\bm \theta)\widetilde{\mathbf W}_k\mathbf h_i(\bm \theta)
		+
		\mathbf h_i^H(\bm \theta)\widetilde{\mathbf V}\mathbf h_i(\bm \theta),
		\label{IR_interference_preserved}
	\end{flalign}
	where the equality follows from
	$\mathbf h_i^H(\bm \theta)\mathbf D_i\mathbf h_i(\bm \theta)=0$. Hence, the SINR of each IR
	remains unchanged, and constraint~\eqref{p3-b} is still satisfied.
	
	For the ER-related information-leakage constraint, since
	$\mathbf D_i\succeq\mathbf 0$, the desired signal power for decoding
	information stream $i$ satisfies
	$\mathbf g_j^H(\bm \theta)\mathbf W_i^\star\mathbf g_j(\bm \theta)
	\leq
	\mathbf g_j^H(\bm \theta)\widetilde{\mathbf W}_i\mathbf g_j(\bm \theta)$.
	Meanwhile, the corresponding interference-plus-AN power satisfies
	\begin{flalign}
		&\sum\nolimits_{k\neq i}
		\mathbf g_j^H(\bm \theta)\mathbf W_k^\star\mathbf g_j(\bm \theta)
		+
		\mathbf g_j^H(\bm \theta)\mathbf V^\star\mathbf g_j(\bm \theta)\notag\\
		&={}
		\sum\nolimits_{k\neq i}
		\mathbf g_j^H(\bm \theta)\widetilde{\mathbf W}_k\mathbf g_j(\bm \theta)
		+
		\mathbf g_j^H(\bm \theta)\widetilde{\mathbf V}\mathbf g_j(\bm \theta)
		+
		\mathbf g_j^H(\bm \theta)\mathbf D_i\mathbf g_j(\bm \theta)\notag\\
		&\geq{}
		\sum\nolimits_{k\neq i}
		\mathbf g_j^H(\bm \theta)\widetilde{\mathbf W}_k\mathbf g_j(\bm \theta)
		+
		\mathbf g_j^H(\bm \theta)\widetilde{\mathbf V}\mathbf g_j(\bm \theta).
		\label{ER_interference_increased}
	\end{flalign}
	Therefore, the eavesdropping SINR at each ER cannot increase, and
	constraint~\eqref{p3-d} remains satisfied.
	
	Thus, the constructed solution
	$\{\mathbf W_i^\star\}_{i=1}^{K_I}$ and $\mathbf V^\star$ is feasible
	to problem~\eqref{p3} and achieves the same objective value as
	$\{\widetilde{\mathbf W}_i\}_{i=1}^{K_I}$ and
	$\widetilde{\mathbf V}$. Therefore, it is also optimal and satisfies
	$\operatorname{rank}(\mathbf W_i^\star)=1$, $\forall i$.
	Theorem~1 is proved.

	\bibliography{Reference}
	\bibliographystyle{IEEEtran}
	
\end{document}